\documentclass[12pt]{article}

\usepackage[top=2cm,bottom=2cm,left=2cm,right=2cm]{geometry}
\usepackage[dvipdfm]{graphicx,color}
\usepackage{amsmath,amsthm,amssymb}
\usepackage{mathrsfs}
\usepackage{enumerate}
\usepackage{fancyhdr,url}

\usepackage{graphicx, subfigure, epsfig}
\usepackage{url}
\usepackage{algorithm}
\usepackage{algorithmic}
\usepackage{amsmath}
\usepackage{amssymb,amsfonts}

\def\u{{\bf u}}

\def\f{{\bf f}}

\def\b{\beta}

\def\g{\gamma}

\def\R{\mathbb R}

\def\p{\partial}

\numberwithin{equation}{section}

\newcommand{\vect}[1]{\mathbf{#1}}

\begin{document}

\title{ Modeling Information Diffusion in Online Social Networks with Partial Differential Equations \thanks{Research supported by NSF Grant CNS-1218212}}

\author{Haiyan Wang, Feng Wang, Kuai Xu\footnote{Email: Haiyan.Wang@asu.edu, fwang25@asu.edu,kuai.xu@asu.edu}\\
School of Mathematical and Natural Sciences\\ Arizona State University, Phoenix, AZ 85069-7100, USA}

\date{}

\maketitle
\begin{abstract}
Online social networks such as Twitter and Facebook have gained tremendous popularity for information exchange. The availability of unprecedented amounts of digital data has accelerated research on information diffusion in online social networks. However, the mechanism of information spreading in online social networks remains elusive due to the complexity of social interactions and rapid change of  online social networks. Much of prior work on information diffusion over  online social networks has based on empirical and statistical approaches. The majority of dynamical models arising from information diffusion over  online social networks involve ordinary differential equations which only depend on time. In a number of recent papers, the authors propose to use partial differential equations(PDEs) to characterize temporal and spatial patterns of information diffusion over online social networks. Built on intuitive cyber-distances such as friendship hops in online social networks, the reaction-diffusion equations take into account influences from various external out-of-network sources, such as the mainstream media, and provide a new analytic framework to study the interplay of structural and topical influences on information diffusion over online social networks. In this survey, we discuss a number of PDE-based models that are validated with real datasets collected from popular online social networks such as Digg and Twitter. Some new developments including the conservation law of information flow in online social networks and information propagation speeds based on traveling wave solutions are presented to solidify the foundation of the PDE models and highlight the new opportunities and challenges for mathematicians as well as computer scientists and researchers in online social networks.
\end{abstract}


\pagenumbering{arabic}
\section{Introduction}

Online social networking has undoubtedly changed the way people communicate and become increasingly popular for information exchange. In recent years, social media (interchangeable with online social networks in this paper) such as Twitter and Facebook has experienced explosive growth. The increasing availability of unprecedented amounts of digital data has accelerated research on information diffusion in online social networks. But the mechanism of information spreading in online social networks remains elusive due to the complexity of social interactions and rapid change of social media.  A better understanding of information diffusion process over social media can effectively predict and coordinate online social activities.  The insight of information spreading process in social media can help increase the efficiency of distributing positive information while reducing unwanted information over social media.

A significant body of research on social media ~\cite{LermanICWSM10, SteegPS11,ChaWOSN08,LermanICWSM10,TangTM11,YangWSM10,Schneider:IMC09,Benevenuto:IMC09,Nazir:IMC09,YuFSKD09} has focused on the measurement and analysis of network structures, user interactions, and traffic characteristics of social media with empirical approaches which utilize data mining and statistical modeling schemes. There is a considerable effort to use mathematical models to understand and predict information diffusion over a time period in online social networks ~\cite{Jin2013,barrat2,Newman2010,Kuma2006,YangLeskovec2010,GhoshWSDM11,Myers2012}. Newman \cite{Newman2003} discussed dynamical processes on complex networks, dynamical models of network growth and dynamical processes taking place on the networks and reports developments on the structure and function of complex networks. Mathematical models based on epidemiological processes have influenced the research on information diffusion \cite{Jin2013,Newman2003}.

However, the deterministic models proposed for online social networks in the literature are largely based on ordinary differential equations(ODEs) which deal with collective social processes over time. Starting from a recent paper \cite{WWX2012}, the authors of this paper proposed to use partial differential equations (PDEs) built on intuitive cyber-distance among online users to study both temporal and spatial patterns of information diffusion process in social media.  One of the basic questions that the models address is that for a given information $m$ initiated from a particular user called {\em source $s$}, the density of influenced users at network distance $x$ from the source at {\em any time $t$} and {\em distance $x$} away from the source $s$.  We validate our models with real datasets collected from popular social media sites, Twitter and Digg. The data-set from Digg consists of millions of votes on top news stories on Digg site during June 2009, and the friendship links among thousands of users who voted during these stories. The experiment results show that the models can achieve over 90 \% accuracy and effectively predict the density of influenced users for a given distance and a given time for a network distance metric with friendship links.

To the best of our knowledge, ~\cite{WWX2012} is the first attempt to propose a PDE-based model for characterizing and predicting the temporal and spatial patterns of information diffusion over social media. According to a recent survey on information diffusion over online social networks by Guille et al. \cite{GuilleSurvey2013}, the PDE model in \cite{WWX2012} is one of the three non-graph based modeling predicative models: epidemiological, Linear Influence Model(LIM) and PDE approaches. The epidemiological models in \cite{GuilleSurvey2013} refer to ODE-based or probabilistic models \cite{Newman2003}.  The LIM approach developed in \cite{YangLeskovec2010} focuses on predicting the temporal dynamics of information diffusion through solving non-negative least squares problems. Our PDE-based models including epidemiological models are spatial dynamical systems that take into account the influence of the underly network structure as well as information contents to predict information diffusion over both temporal and spatial dimensions.

The PDE-based models we developed directly address a number of concerns in studying information diffusion in online social networks with epidemiological models.  Tufekci et al. \cite{Tufekci2013} observed that there are significant differences between information traveling in social media and the spreading of germs in that online users are exposed to information from a wide range of sources and not only from the networks they are connected to.  The same issue also was raised by Myers et al. \cite{Myers2012} (also see \cite{GuilleSurvey2013} ) where two different diffusion processes, internal and external influence, were discussed.  The internal influence results from the structure of the underlying network; the external influence comes from various out-of-network sources, such as the mainstream media. It is estimated in \cite{Myers2012} that almost 27\% of information volume in Twitter can be attributed to the external influence. \cite{Myers2012} noticed that nearly all epidemiological models for online social networks only focus on the internal influence, while neglecting the external influence. However, the probabilistic model in \cite{Myers2012} primarily focuses on separating the external influence from the internal influence, and quantifying the impact of the external influences on information adoption over time. The PDE-based models we developed integrate the effect of both the structured-based process (internal influence) and content-based process (external influence) through dynamical systems in both temporal and spatial dimensions. It is plausible to see that the network of social relationships and the set structure of topical affiliations form the backbone of online social media(Romero et al. \cite{Romero2013}) and the popularity of the content of information is the key driving force behind the external influence. As such, our PDE models provide a new analytic framework towards a better understanding of  information diffusion  mechanisms  by studying the interplay of structural and topical influences.

Our work extends the applications of PDEs into the research of information diffusion in online social network. In the last few decades, there have occurred numerous new developments in mathematical analysis of reaction-diffusion systems. In this paper, in addition to a review of a number of recent PDE models for information diffusion in our recent papers, some new developments including the conservation law for information flow in social media are presented to provide a more rigorous justification for the PDE models. We discuss stability, bifurcation, free boundary value problem, information propagation speeds based on analysis of traveling wave solutions for interaction models. The theoretical advances in partial differential equations can provide an analytic tool to reveal mechanisms of information diffusion. For example, analysis of the free boundary value problem arising from social media in Section \ref{mathh} leads to a simple formula for how fast information is traveling. Surprisingly, the formula is almost the same as the celebrated result of Fisher, Kolmogorov, Petrovsky and Piscounov  in 1937 ~\cite{Fisher,Kolmogorov} on the spreading of advantageous genes. These results provide reasonable predications for how parameters influence information diffusion over social media.  However, because of the complexity of human interactions and rapid change of social media, PDE models from social media can be quite complex and difficult to study analytically. The short survey presents a number of simple PDE models arising from social media and highlights the new opportunity and challenge for mathematicians as well as computer scientists and researchers in social media.

This paper is organized as follows. Section \ref{spatial-tep} discusses the spatial-temporal phenomena in social media. Section \ref{conserv} introduces the conservation law of information flow in online social networks. Section \ref{models} presents a number of PDE-based models to describe information flow and validations of the models with real datasets.  Section \ref{interaction} examines several complex spatial models for complex interactions in social media. Section \ref{mathh} discuss a number of related mathematical problems and gives some theoretical results for the problems.  Section \ref{endd} concludes the paper with a wide range of challenges in modeling online social networks.

\section{Information Diffusion over Social Media}\label{spatial-tep}
\subsection{Digg and Twitter Data}
In order to develop and validate PDE-based models, we use real datasets collected from Twitter.com, the largest micro-blogging site, and Digg.com, the most popular news aggregation sit. In Digg, registered users can post links of news stories and blogs to Digg.com.  Other registered users can vote and comment on the submitted news links. Digg users can connect one to another by establishing friendship relationship called ``follow''.  The initiator or source of a news link is the voter who first posts the news to the Digg site. In addition to followers, who can view and choose to vote the news submitted by the friend he/she follows, Digg users, who do not friend with the initiator directly or indirectly, will also be able to view and vote the news once news is promoted to the front page after certain time.  A user can also search for particular news at the web site and vote for it. The news propagation that does not result from the structure of the online social networks behaves somewhat randomly, which resembles random walk in the development of partial differential equations. Thus the Digg data provides a very good opportunity for us to study the impact of the friendship relationships on the process of information spreading with partial differential equations.

We will validate our PDE models with the data-sets from Digg consisting of the $3553$ news stories that are {\em voted} (also called {\em digged}) and promoted to the front page of {\tt www.digg.com} due to the popularity during June 2009. In total, there are more than 3 millions votes cast on these news stories from over $139,409$ Digg users. In addition, the data-sets also include the directed {\em friendship} links among the Digg users who have voted these news stories. Based on these friendship links, we construct a directed social network graph among these Digg users. For each of the news stories, the dataset includes the user id of all the voters during the collection period, and the timestamps when votes are cast.

We also collected data from Twitter. Twitter has much in common with Digg. Within the twitter social network, users "follow" other people with twitter accounts. These users can follow their friends, celebrities, or even famous politicians. By being a "follower", one can view the tweets, and also, "retweet" a person's message. When a person "retweets" a status or a picture, he or she is reposting the tweet so that his or her followers can now view the tweet. By retweeting, followers are practicing information diffusion through online social networking.  The time stamp and the social network graph give us the opportunity to study the temporal and spatial patterns of information propagation.

\subsection{ Cyber-distance Based on Friendship Hops}

\begin{figure}[htbp]
    \centering{
    \includegraphics[width=2.5in]{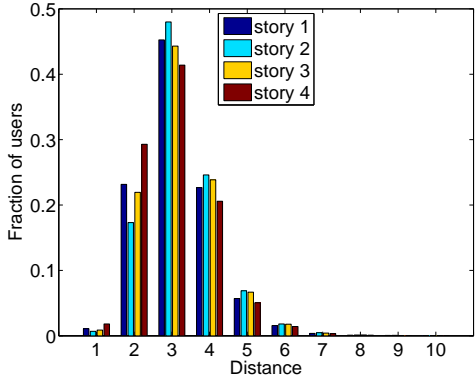}}
 \caption{Distribution of neighbors of four stories}
\label{fneighbor}
\end{figure}

In online social networks, cyber-distances or friendship hops play a significant role in information diffusion. Cyber-distance or social distance is used to measure the closeness of users in online social networks. An intuitive approach for defining the distance between two users is to use the number of friendship links in the shortest path from one user to another in the social network graph. We use {\em friendship hops} to refer to the number of friendships links or hops. Thus  the distance between the initiator and any other user is defined as the length (the number of friendship hop) of the shorted path from the initiator to this user in the social network graph \cite{WWX2012}. Clearly, the direct followers of the initiator have a distance of $1$, while their own direct followers have a distance of $2$ from the initiator, and so on. Figure~\ref{fneighbor} shows the distance distributions of the direct and indirect followers from  Digg users who have initiated one
or more top news stories in the Digg dataset. As we can see from the figure, the majority of online social network users have a distance of $2$ to $5$ from the initiators. In this figure, for all four stories, the distance $3$ users accounts for more than $40\%$ of all the users from the initiator directly or through other users. As the
distance increases from $6$ to $8$, the number of social networks users reachable from the initiator drops dramatically.

To be more precisely, let $U$ denote the user population in an online social network, and
$s$ is the source of information such as a news story that
starts to spread in social media. Based on the distance from social network users from this source,
the user population $U$ can be divided into a set of
groups, i.e., $U = \{U_1, U_2, ... U_i, ..., U_m\}$, where $m$ is the maximum distance
from the users to the source $s$. The group $U_x$ consists of users that share the same
distance of $x$ to the source.

While the social distance in this paper is based on friendship hops, its definition of cyber-distance can be flexible and can be defined as other measurements. For example, Section \ref{interest} discusses an alternative way to define distance metrics based on shared interests.

\subsection{Temporal and Spatial Patterns of Information Diffusion}

In order to model the information diffusion process in temporal and spatial dimensions, we examine the real datasets to analyze information spreading patterns over social media. With the definition of distance, all users can be divided into distance groups based on their distance from the news submitter.  As a news story propagates through the Digg or Twitter network, users express their interests in the news by voting or retweeting for it. We call such users as {\em influenced users} of the information.

\begin{figure*}[htbp]
  \centering
  \subfigure[Density of influenced users of s1]{
    \includegraphics[width=1.6in]{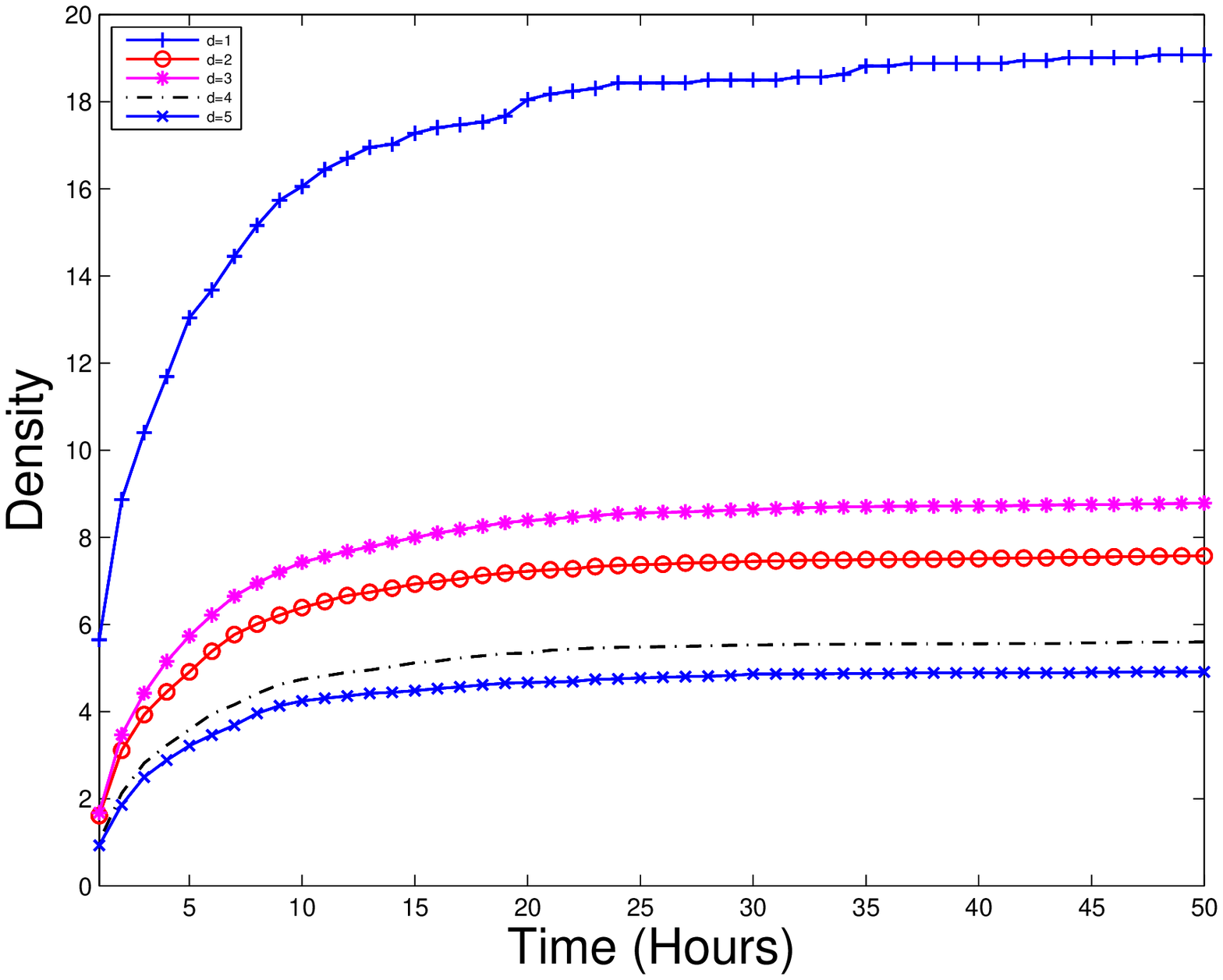}}
  \subfigure[Density of influenced users of s2]{
    \includegraphics[width=1.6in]{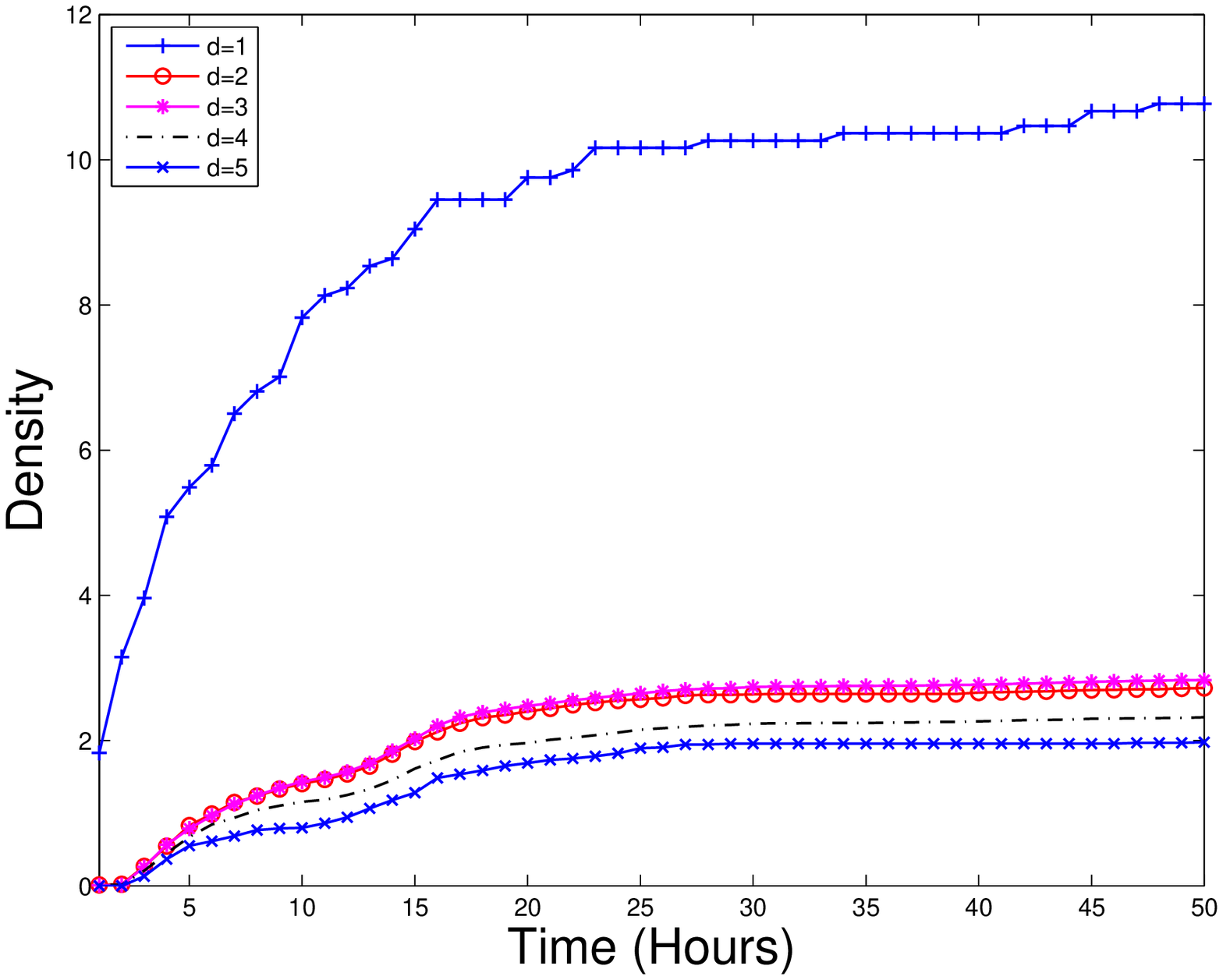}}
  \subfigure[Density of influenced users of s3]{
    \includegraphics[width=1.6in]{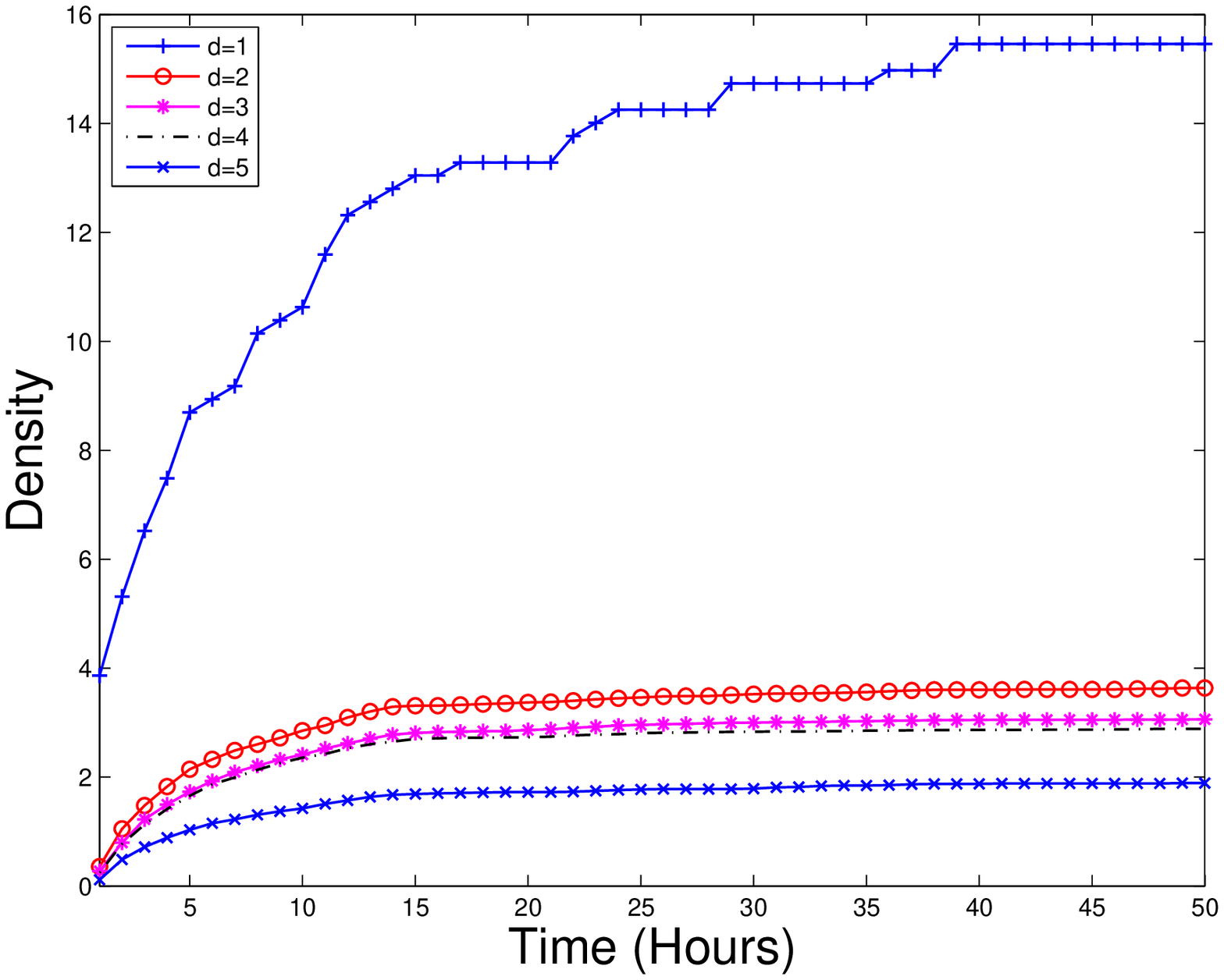}}
  \subfigure[Density of influenced users of s4]{
    \includegraphics[width=1.6in]{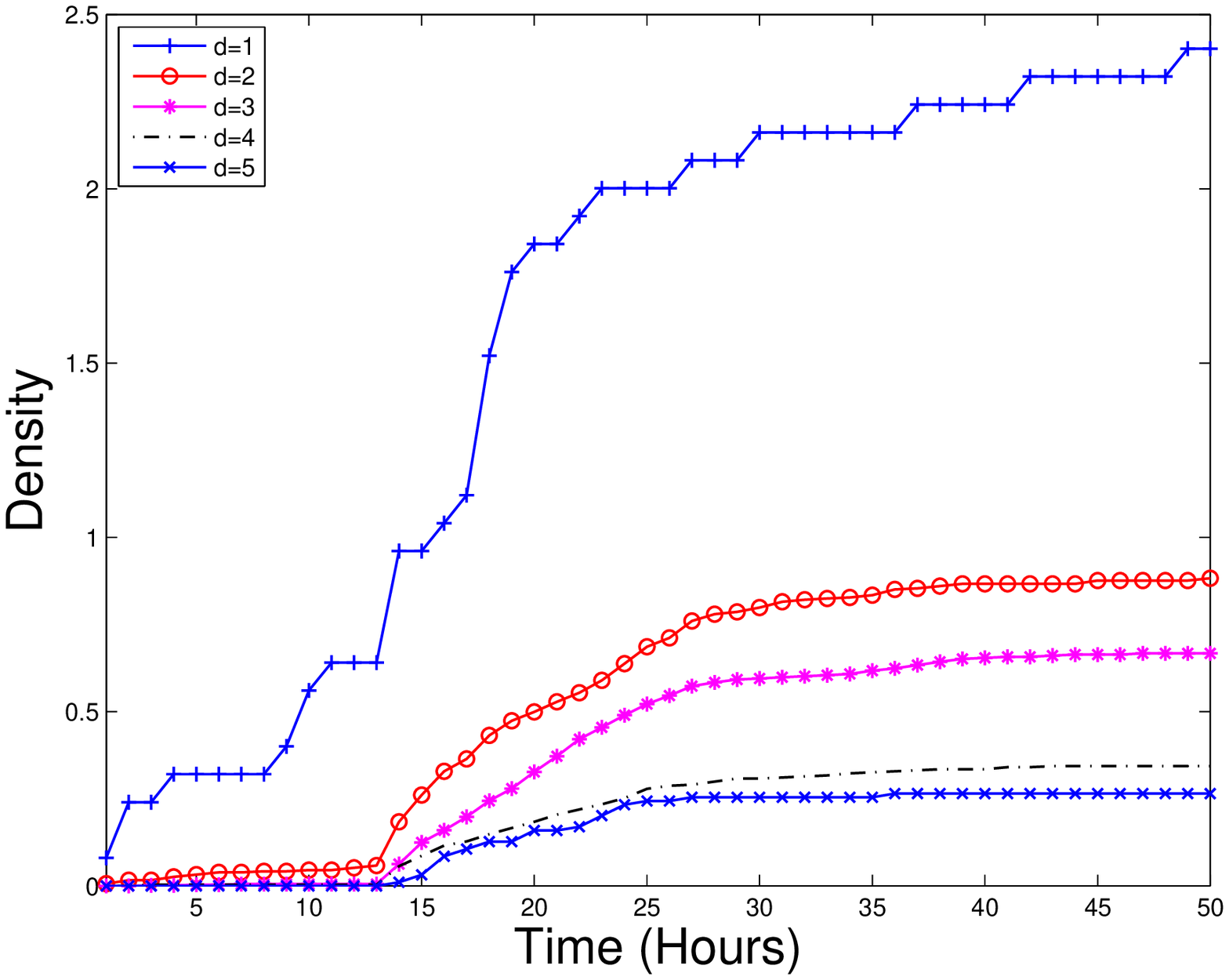}}
 \caption{Densities of influenced users over 50 hours for s1 to s4}\label{fdensity}
\end{figure*}

\cite{WWX2013} studies the impact of friendship hop distance on the information diffusion process by measuring the
{\em density of influenced users} within the same distance, which is the percentage of
influenced users over the total number of users within a given distance. In the context
of Digg social networks, we consider the users who have voted the news story as
influenced users. Figure~\ref{fdensity}[a-d] illustrate the density of influenced users
(with the distances of $1-5$) over the initial 50 hours since the news stories was posted
on Digg for four example news stories, respectively. Each curve in Figure~\ref{fdensity}[a-d] represents the density at a different distance.

We can observe from Figure~\ref{fdensity}[a-d] that the densities of influenced users at different distances show consistent evolving patterns rather than increasing or decreasing with random fluctuations. The temporal and spatial patterns resemble dynamics of evolution equations involving both time and space variables.

In addition, \cite{WWX2013} validates the observations for all news stories in the Digg data set.  It is concluded that $94.9\%$ of all news stories have the similar consistent evolving patterns. For most of the news stories, densities of influenced users decrease as the distances of the users increase, reconfirming that friendship is an important channel of information spreading.  Therefore, mathematical models, in particular, evolution equations involving both time and space variables,  can be used to describe the evolution dynamics of information diffusion over social media.

It is worth noting that there are some differences between information diffusion in online social networks and spatial biological process in mathematical biology. In spatial ecology, the diffusion process often refers to the fact that animals  move randomly from one physical location to another. In the context of online social network, online users simply pass on information from one to another and do not necessarily change their network distances within the lifetime of the information.

\section{Conservation Law of Information Flow}\label{conserv}


\subsection{Embedment of Diffusion Process into Euclidean space}

Based on cyber-distance from a source, one could breakdown the user population $U$ into a set of
groups, i.e., $U = \{U_1, U_2, ... U_i, ..., U_m\}$, where $m$ is the maximum distance
from the users to the source $s$. The group $U_x$ consists of users that share the same
distance of $x$ to the source.  Following \cite{WWX2012},  we use the $x$-axis as the social distance and embed the density $U_x$ at the location $x$. Let $I(x,t)$ denote the density of influenced users at distance $x$ and time $t$. For Digg news, $I(x,t)$ is the ratio of the number of influenced users with a distance of $x$ at time $t$ over the total number users in $U_x$. For Twitter news, $I(x,t)$ is the number of influenced users with a distance of $x$ in $U_x$ at time $t$ as Twitter has huge numbers of registered users (500 million registered users in 2012  \cite{wikit}) and the ratio for Twitter would be too small to study.

\begin{figure}[htbp]
    \centering{
    \includegraphics[width=3.0in]{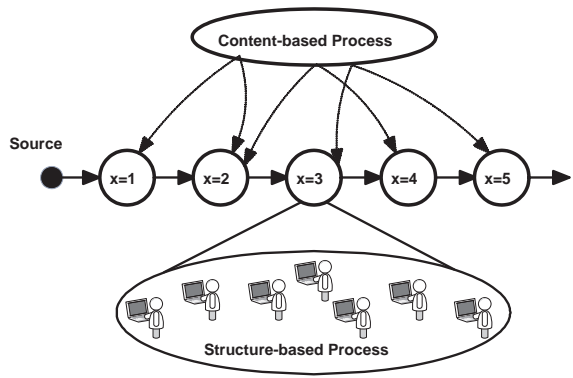}}
 \caption{Information diffusion process in online social networks}
\label{fprocess}
\end{figure}

As information propagates over social media, users promote the information through retweeting, commenting, searching, voting, forwarding and other activities.  In general, two decisive components for information diffusion in online social networks are the graph structure of social networks( follower graphs) and the content of the information, which form the backbone of online social networks \cite{Romero2013}. Online users are subject to information from a wide range of sources, not just those networks they are connected to  ~\cite{Tufekci2013}. In our setting, because group $U_x$ consists of users from the same social distance from a source, the growth of the influenced users within the group may be viewed as a result of the network structure. Other activities to promote information diffusion such as search do not result from the network structure and may happen randomly for various reasons, in most cases, mainly because of the content of the information.

As such, we divide the information diffusion process in online social networks into two separate processes, structure-based process and content-based process. The content-based and structure-based processes in Fig. \ref{fprocess} resemble the external and internal influences, respectively, in online social networks in \cite{Myers2012}. The interplay of the two processes essentially accounts for the change of the density of influenced users $I(x,t)$. The structure-based process represents the information spreading among users in $U_x$ with the same distance because of their direct links to those who already are already influenced.  The content-based process measures information spread among users at different distance due to various other activities that result from the popularity of content of the information.  The content-based process is usually bidirectional or reciprocal in a manner of random walk.  Figure~\ref{fprocess} conceptually illustrates the interplay of the two processes in an online social network.  The content-based and structure-based processes are named slightly different in our previous papers \cite{WWX2012,LZW1013}, but refer to the same processes in the context of online social networks.

As social media rapidly gains worldwide popularity in recent years,  many social media sites experience an explosive growth of registered online users. For example, Twitter has 500 million registered users in 2012. This gives rise to extremely complex and large network graphs in online social networks.  If we introduce a slightly more complex distance metric from its underlying network topology, the number of the subsets in $U$ can increase dramatically. Therefore the user population $U$ will be embedded to more dense points in some interval on the $x$-axis. In particular, when we discuss traveling wave solutions, it is assumed that these discrete points are enough dense on a large section of the $x$-axis that can be mapped to $(-\infty, \infty)$.

\subsection{Formulation of Conservation Law of Information Flow}

Conservation laws or basic balance laws play a crucial role in the development of partial differential equation models in Physics, Mathematical Biology and other fields. A conservation law is a mathematical formulation of the basic fact that
the rate at which a given quantity  changes in a given domain must equal the rate
at which it flows across its boundary plus the rate at which
it is created, or destroyed, within the domain. Once we embed the information propagation process into Euclidean spaces, the formulation of the conservation law for information flow is similar to that for spatial biology ~\cite{Murray1989}. We emphasize differences and their interpretations in social media.

In social media, the quantity is the amount of information spreading such as the density of influenced uses, denoted by $I =I(x, t)$ and measured in amount per unit length along the $x$-axis since we are embedding the users of entire network into a one-dimensional space.  We assume that any change in the amount of information be restricted to one spatial dimensional tube where each cross-section is labeled as the spatial variable $x$.  While only discrete set of points ($U_x$) in the $x$-axis which is meaningful for social media, we can extend the discrete points into a continuous interval. With this understanding, we can derive the conservation law of information flow.  Our spatial models are direct applications of the conservation law of information flow, and will be validated by real data sets.

For simplicity, we assume that a constant $A$ is the cross-sectional area of the tube. Thus the amount of information in a small section of width $dx$ is $I(x, t)Adx$. Further, we let $J =J(x, t)$ denote the flux of the quantity at $x$, at time $t$. The flux measures the amount of the quantity crossing the section at $x$ at time $t$, and its units are given in amount per unit area, per unit time.  In social media, $J$ reflects the content-based diffusion process in Fig. (\ref{fprocess}) and does not result from the structure of the underlying network.   By convention, flux is positive if the flow is to the right, and negative if the flow is to the left.

In social media, influenced users in $U_x$ may increase because they directly link or follow those who are already influenced.  Let $f = f(I,x, t)$ denote the given rate at which the information is created within the section at $x$
at time $t$.  $f$  represents the structure-based process in Fig. (\ref{fprocess}) and is a result of local growth due to the underlying network structure.  The structure-based diffusion process has much in common with the internal influence in \cite{Myers2012}.  $f$ can be negative in social media if some kind of deletion occurs.  $f$ is measured in amount per unit volume per unit time. In this way, $f(I,x, t)Adx$ represents the amount of information that is created in a small width $dx$ per unit

We now can formulate the law by considering a fixed, but arbitrary, section $a \leq x \leq b$ of the domain.  The rate of change of the
total amount of the information in the section must equal to the rate at which
it flows in at $x = a$, minus the rate at which it flows out at $x = b$, plus the
rate at which it is created within $a < x < b$. In mathematical formulation, for any section $a \leq x \leq b$, $$
\frac{d}{dt}\int^b_aI(x,t)Adx=AJ(a,t)-AJ(b,t)+\int_a^bf(I,x,t)Adx
$$
From the fundamental theorem of calculus, $
J(a,t)-J(b,t)=-\int_a^b \frac{ \partial J}{\partial x} dx
$. Because A is constant, it may be canceled from the formula. We arrive at, for any section $a \leq x \leq b$, $$
\int^b_a\big(\frac{ \partial I}{\partial t}+\frac{ \partial J}{\partial x}-f(I,x,t)\big)dx=0
$$
It follows that the fundamental conservation law of information flow is $$
\frac{ \partial I}{\partial t}+\frac{ \partial J}{\partial x}=f(I,x,t)
$$

$J$  does not necessarily result from direct social links and behaves like random walk.  For example, in Digg network, besides the fact that a follower votes for news posted by its followee, a user can also vote for any news that he/she is interested in while the news is promoted to the front page, or through search engines provided by the network. In Twitter,   the symbol \# followed by a few characters, called a hashtag, is used to mark keywords or topics in a tweet.  With the hashtag symbol anyone can search for the set of tweets that contain a hashtag.  It is estimated that Twitter handles 1.6 billion search queries per day \cite{wikit}. The use of hashtags increases propagation of tweets.  Also Twitter users can send $@$-messages publicly to a specific user by including the "\@" character before the receiving person's username in their tweet.  This unstructured phenomenon ``jumps" across the network and appears at a seemingly random node \cite{Myers2012}.  The action results from the relevance of the content of information rather than the structure of the follower graph of a network. In general, information flows from high density to low density and therefore a simple expression of flux $J$ can be
\begin{equation}\label{eq1f}
J= -d \frac{ \partial I}{\partial x}
\end{equation}
which results from a principle analogous to Fick's law(\cite{Murray1989}) in Biology or Physics. The minus sign describes the flow is down the gradient. $d$ represents the popularity of information which promotes the spread of the information through non-structure based activities such as search. For now $d$ can be viewed as an average and therefore is a constant. In general, it may  be dependent on $u,x, t$, which we will investigate in the paper.  Now we obtain the following  PDE model to describe information flow.
\begin{equation}\label{eq10}
\frac{ \partial I}{\partial t}= d \frac{\partial^2 I}{\partial x^2}+f(I,x,t)
\end{equation}

\begin{table}[h]
\begin{center}
\begin{tabular}{|c|c|c|}
  \hline
   Symbol& Description  \\
  \hline
  $d \frac{\partial^2 I}{\partial x^2}$ &  \shortstack{Diffusion process \\ (random walk)}  \\
  \hline
  $f(I,x,t)$ & \shortstack{ Local growth \\( birth and death)} \\
  \hline

\end{tabular}
\caption{ Equation (\ref{eq10}) in mathematical biology with physical distance}
\label{comparsionBIO}
\end{center}
\end{table}

\begin{table}[h]
\begin{center}
\begin{tabular}{|c|c|c|c|}
  \hline
   Symbol &  Description   & \shortstack{  Similar concept \\in the literature }& \shortstack{Key reason to \\view the information}\\
  \hline
  $d \frac{\partial^2 I}{\partial x^2}$ & \shortstack{ Content-based \\ (random action) } & \shortstack{ External influence \cite{Myers2012}  \\ (various webpages \\such as cnn.com)}& \shortstack{ Search  or others\\ (due to the popularity \\of content)}\\
  \hline
  $f(I,x,t)$ & \shortstack{Structure-based \\ (structured action)} &  \shortstack{ Internal influence \cite{Myers2012} \\ (diffusion over the edges\\ of the network)} & Follower graph \\
  \hline

\end{tabular}
\caption{Equation (\ref{eq10}) in online social media with friendship hops as distance in Fig. \ref{fprocess}}
\label{comparsionOSN}
\end{center}
\end{table}

Tables \ref{comparsionBIO} and \ref{comparsionOSN} compare the difference of the interpretations of PDE models in both mathematical biology and online social networks in the setting of Fig. \ref{fprocess}.  The structure-based process can be viewed as the growth of population due to local growth in mathematical biology.  The content-based process is similar to the diffusion process in mathematical biology and behaves in a manner of random walk.   The content-based and structure-based processes in Fig. \ref{fprocess} resemble the external and internal influences, respectively, in online social networks in \cite{Myers2012}. The key difference of the structure-based and content-based processes is that the former results from information received from the follower graph; the latter results from information received from search or other actions because of the popularity of content rather than the follower graph.

\section{Partial Differential Equation Models}\label{models}
In this section, we will discuss three Reaction-Diffusion equation models which will be validated by real datasets. Two of them have been published in \cite{WWX2012} and \cite{WWX2013}. Spatial models have been used in mathematical biology, sociology, economics, and physics to model spatial-temporal patterns.  Spatial models are able to provide a quantitative way to integrate local data about interactions between individual users into global conclusions about news spread over social media. By introducing PDEs into the context of social media, we capture the similarity and difference between spreading of epidemics in biology and the information spreading in online social networks.

\subsection{Diffusive Logistic Model}

In this subsection we review the diffusive logistic model in the authors ~\cite{WWX2012} for characterizing the temporal and spatial patterns of
information cascading over social media.  The model shall be validated with the data set from Digg. Logistic model is believed to be the simplest nonlinear model to capture the population dynamics where the rate of reproduction is proportional to both the existing population and the amount
of available resources ~\cite{Murray1989}.  It has been widely used to describe various
population dynamics and predict growth of bacteria and tumors
over time ~\cite{Murray1989}. The structure-based process in Fig. (\ref{fprocess}) is modeled with a simple nonlinear equation. Logistic equation is defined as follows.
Denoting with $N$ the population at time $t$, $r$ the intrinsic
growth rate and $K$ the carrying capacity which gives the upper
bound of $N$, the population dynamics are governed by:
\begin{equation}
\frac{d N}{dt} = rN(1-\frac{N}{K})
\end{equation}
where $\frac{d N}{dt}$ is the first derivative of $N$ with respect to $t$. In the context of online social networks, the term $rN(1-\frac{N}{K})$ describes the impact of the network structure on the growth of $I(x,t)$, the density of influenced users at the
distance $x$ during time $t$.

\begin{figure}[ht]
  \centering{
    \includegraphics[width=2.0in]{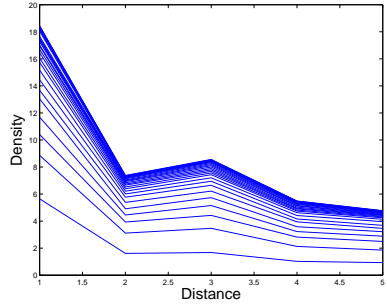}
  }
 \caption{Density of influenced users over 50 hours with friendship hops as distance} \label{f332346.hop.distance}
\end{figure}

$r$ reflects the decay of news influence with respect to time $t$. While some information can take a longer period of time to spread in social media \cite{Cha2009}, news diffusion in social media is time-sensitive and the influence of news stories decays drastically as time elapses.  Figure~\ref{f332346.hop.distance} illustrates the spread of the most popular story in the digg dataset in the temporal perspective. It shows that interests in news decay exponentially over time. The x-axis is the distance, y-axis is the density of the influenced users, each line represents the density at time $t$ where $t$ is 1 hour, 2 hour and up to 50 hours after the submission of the initial news. The gap of density decreases at time pass by. From our experiments exponential functions of decay seems plausible for modeling the rapid decay of news with respect to time.

It is conceivable that the rate of influence of a news decays experientially, which can also be observed in Fig. \ref{f332346.hop.distance}. The decay process can be modeled by the following ordinary differential equation
\begin{equation}\label{ode}
\begin{split}
&\frac{d r(t)}{d t} = -\alpha r(t)+\beta\\
& r(1) = \gamma \\
\end{split}
\end{equation}
where $\frac{d r(t)}{d t}$ is the rate of change of $r$ with respect to time $t$,  $\alpha$ is the decay rate, $\gamma$ is the initial rate of influence. $\beta$ represents the residual rate as time increase, which can be very small.  Solving for $r$ in (\ref{ode}), we obtain
\begin{equation}\label{rt}
r(t)=\frac{\beta}{\alpha}-e^{-\alpha (t-1)}(\frac{\beta}{\alpha}-\gamma)
\end{equation}

Base on the conservation law of information flow, combining the structure-based  process and content-based process together gives the following diffusive logistic equation:
\begin{equation}\label{eq1}
\begin{split}
&\frac{\partial I}{\partial t}=d \frac{\partial^2 I}{\partial x^2}+r I(1-\frac{I}{K})\\
& I(x, 1) = \phi(x), \;\;l\leq x \leq L  \\
&\frac{ \partial I}{\partial x}(l,t)=\frac{ \partial I}{\partial x}(L,t)=0, \;\;  t \geq 1 \\
\end{split}
\end{equation}

where
\begin{itemize}
\item $I$ represents the density of influenced users with a distance of $x$ at time $t$;
\item $d$ represents the popularity of information which promotes the spread of the information through  non-structure based activities such as search;
\item $r$ represents the intrinsic growth rate of influenced users with the same distance, and measures how fast the information spreads within the users with the same distance;
\item $K$ represents the carrying capacity, which is the maximum possible density of influenced users at a given distance;
\item $L$ and $l$ represent the lower and upper bounds of the distances between the source $s$ and other social network users;
\item $\phi(x) \geq 0$ is the initial density function, which can constructed from history data of information spreading. Each information has its own unique initial function;
\item $\frac{ \partial I}{\partial t}$ represents the first derivative of $I$ with respect to time $t$;
\item $\frac{ \partial^2 I}{\partial x^2}$ represents the second derivative of $I$ with respect to distance $x$;
\end{itemize}

$\frac{ \partial I}{\partial x}(l,t)=\frac{ \partial I}{\partial x}(L,t)=0$ is the Neumann boundary condition~\cite{Murray1989},
which means no flux of information across the boundaries at $x=l,L$. This assumption is plausible for social media since the users cluster in a number of groups $U_x$. We also assume $\phi(x) \geq 0$ is not identical to zero and the maximum principle implies that (\ref{eq1}) has a unique positive solution $I(x,t)$ and $0 \leq I(x,t) \leq K$.

\subsubsection{Initial Density Function Construction}

In general, we assume that the initial density function is given and can be constructed using the data collected from the initial stage of information diffusion. Specifically, $\phi$ is a function of distance $x$ which captures the density of influenced user at distance $x$ at the initial time when a news story is submitted. In online social networks, it is only possible to observe discrete values for the initial density function because the distance $x$ is discrete.  The initial density is the influenced user distribution when time $t=1$. As in \cite{WWX2012}, we apply an effective mechanism available in Matlab cubic spline package, called {\em cubic splines
interpolation}~\cite{Gerald1994}, to interpolate the initial discrete data in constructing $\phi(x)$.
Using this process, a series of unique cubic polynomials
are fitted between each of the data points, with the stipulation that the obtained curve
is continuous and smooth. Hence $\phi(x)$ constructed by the cubic splines
interpolation is a piecewise-defined function and twice continuous differentiable.
After cubic splines interpolation, we simply set the two ends to be flat to satisfy
the second requirement since in this way the slopes of the density function $\phi(x)$
at the left and right ends are zero.

\subsubsection{Accuracy of Diffusive Logistic Model} \label{accusdlm}

In this subsection, we evaluate the performance of the proposed linear diffusive model by comparing the density calculated by the model with the actual observations in the Digg data set. We first present the model accuracy for the most popular news story. Accuracy of the predicated value of a model against an actual value is defined as follows.

\begin{equation}\label{accuracy}
\text{model\_accuracy} = 1-\frac{|\text{predicted\_value} - \text{actual\_value}|}{\text{actual\_value}}
\end{equation}

\begin{figure}[ht]
  \centering{
    \includegraphics[width=3.0in]{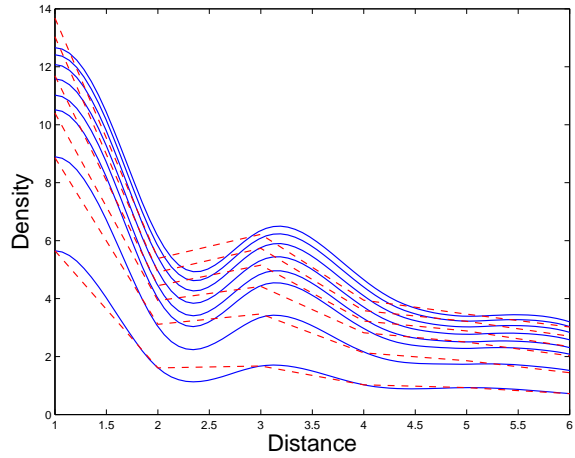}
  }
 \caption{Prediction of (\ref{eq1})  vs. real data of story 1 with 24099 votes}
 \label{f332346.match}
\end{figure}

We numerically solved the model with Matlab. Figure~\ref{f332346.match} illustrates the predicting results for an example news story (story $1$)
with the proposed model, where the $x$-axis is the distance measured by {\em friendship hops}, while
the $y$-axis represents the density of influenced users within each distance. The solid
lines denote the {\em actual} observations for the density of  influenced users
for a variety of time periods (i.e., 1-hour, 2-hours, 3-hours, 4-hours and 5-hours),
while the dashed lines illustrate the {\em predicted} density of influenced
users by the model. As we can see,  the proposed model is able to accurately predict the density
of influenced users with different distance over time. The values of
the two parameters $K$ and $d$ in this case are $25$, and $0.01$. $r(t) = 1.4e^{-1.5(t-1)} + 0.25$.
Table~\ref{tab-prediction} gives the numerical value of Figure~\ref{f332346.match}. It is clear that the model has high precision in terms of prediction.

\begin{table}[htbp]
\centering
\scriptsize
\begin{tabular}{|c|c|c|c|c|c|c|} \hline

Distance & Average & t = 2 & t = 3 & t = 4 & t = 5 & t = 6 \\ \hline

1 & 98.27\% & 97.47\% & 97.74\% & 97.48\% & 99.55\% & 99.09\% \\ \hline

2 & 86.99\% & 93.59\% & 96.63\% & 87.16\% & 80.80\% & 76.78\%  \\ \hline

3 & 90.28\% & 83.23 \% & 87.98\% & 90.99\% & 93.35\% & 95.94\%  \\ \hline

4 & 92.98\% & 86.75\% & 91.39\% & 99.00\% & 95.68\% & 92.06\%  \\ \hline

5 & 93.77\% & 89.05\% & 91.61\% & 97.79\% & 97.92\% & 92.49\%  \\ \hline

6 & 94.56\% & 90.03\% & 89.48\% & 96.04\% & 97.57\% & 99.67\%  \\ \hline

\end{tabular}

\caption{Prediction accuracy of Accuracy of (\ref{eq1}) with friendship hop as distances for story $s1$}
\label{tab-prediction}
\end{table}

\subsection{Linear Diffusive Model}
The logistic growth model in \cite{WWX2012} is the first spatial model to account for the phenomenon that the initial stage of the increase of influenced users is approximately exponential; then, as saturation begins, the growth slows, and eventually, growth stops. It can achieve a high accuracy as we discuss in the last subsection. In this subsection, we present a more simple
linear function to model the growth of influenced users in online social networks \cite{WWX2013} by the authors, Wu and Xia.   The linear model takes into account the effects of heterogeneity in cyber-distance and news decay with respect to time.  As indicated in Fig. \ref{fneighbor}, the distribution of the density of influenced users in distance is not homogeneous. The majority of users are in the groups with distances $3$ and $4$. This heterogeneity in distance leads to the assumption that the growth function is dependent on location $x$. The concavity of the shape of Fig. \ref{fneighbor} further suggests that
we can use the following concave down quadratic function $h(x)$ to describe this heterogeneity in distance.
\begin{equation}\label{hx}
h(x)=-(x-\rho)(x-\sigma)
\end{equation}
The coefficient of $x^2$ in $h(x)$ is scaled to be $-1$. $h(x)$ reflects the rate of the change of influenced users with respect to distance $x$.  The simplest way to model the growth of influenced users as linear function of $I$.  Let
$$
f=r(t)h(x)I
$$
We can think of $r(t)$ as the average of all distances, and likewise, $h(x)$ as the average of all times.  Thus, combining the structure-based process  (\ref{fprocess}) and the growth process together, the fundamental law of information flow gives the following the linear diffusive equation
\begin{equation}\label{eq2}
\begin{split}
&\frac{\partial I}{\partial t}=d \frac{\partial^2 I}{\partial x^2}+r(t)h(x)I\\
& I(x, 1) = \phi(x), \;\;l< x < L  \\
&\frac{ \partial I}{\partial x}(l,t)=\frac{ \partial I}{\partial x}(L,t)=0, \;\;  t > 1 \\
\end{split}
\end{equation}

\subsubsection{Accuracy of Linear Model }
\begin{figure}[h]
  \centering
  \subfigure[Predicted (blue, solid) vs. Actual data (red, dotted)]{
    \includegraphics[width=2.5in]{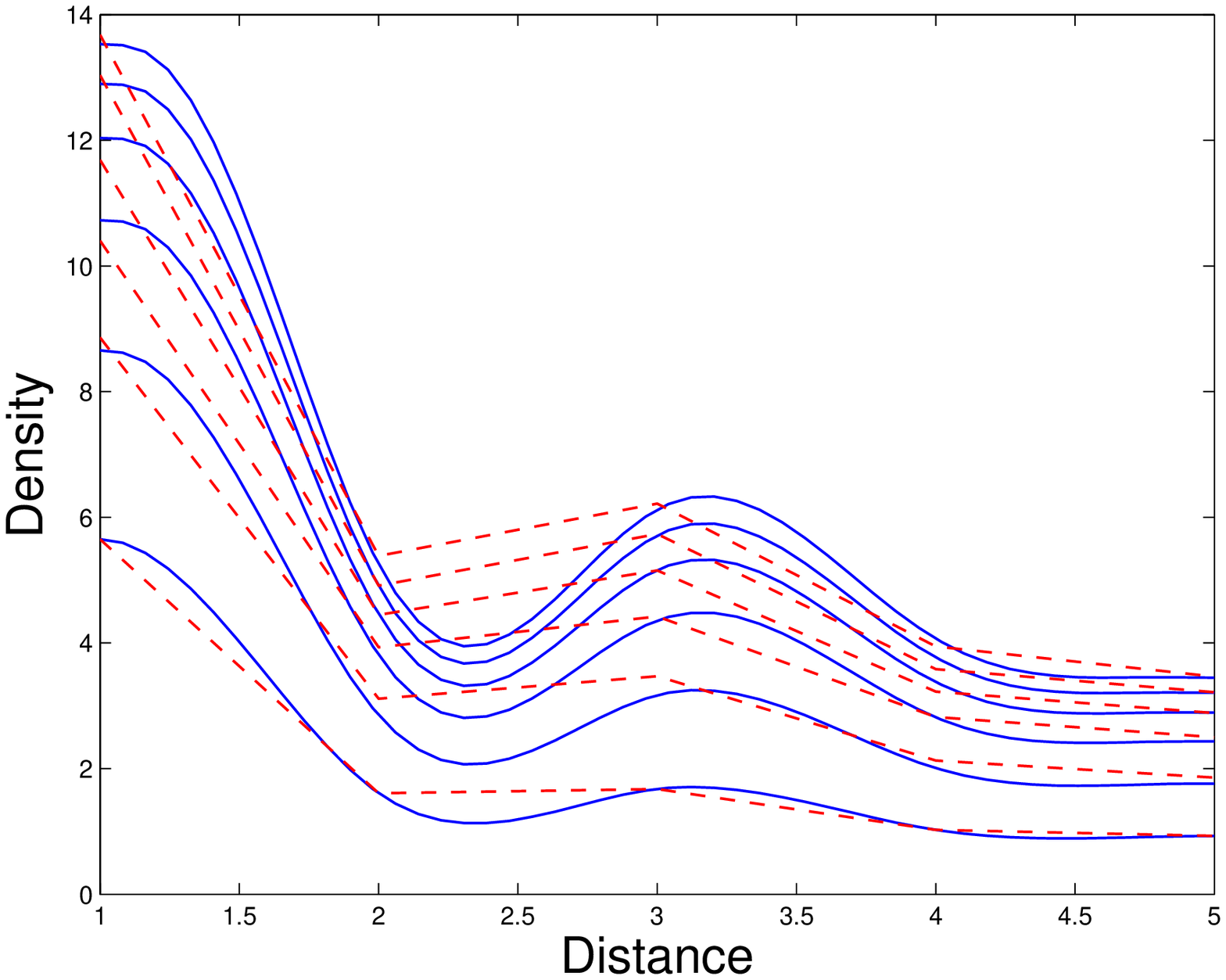}}
  \subfigure[$r(t)$ ]{
    \includegraphics[width=1.6in]{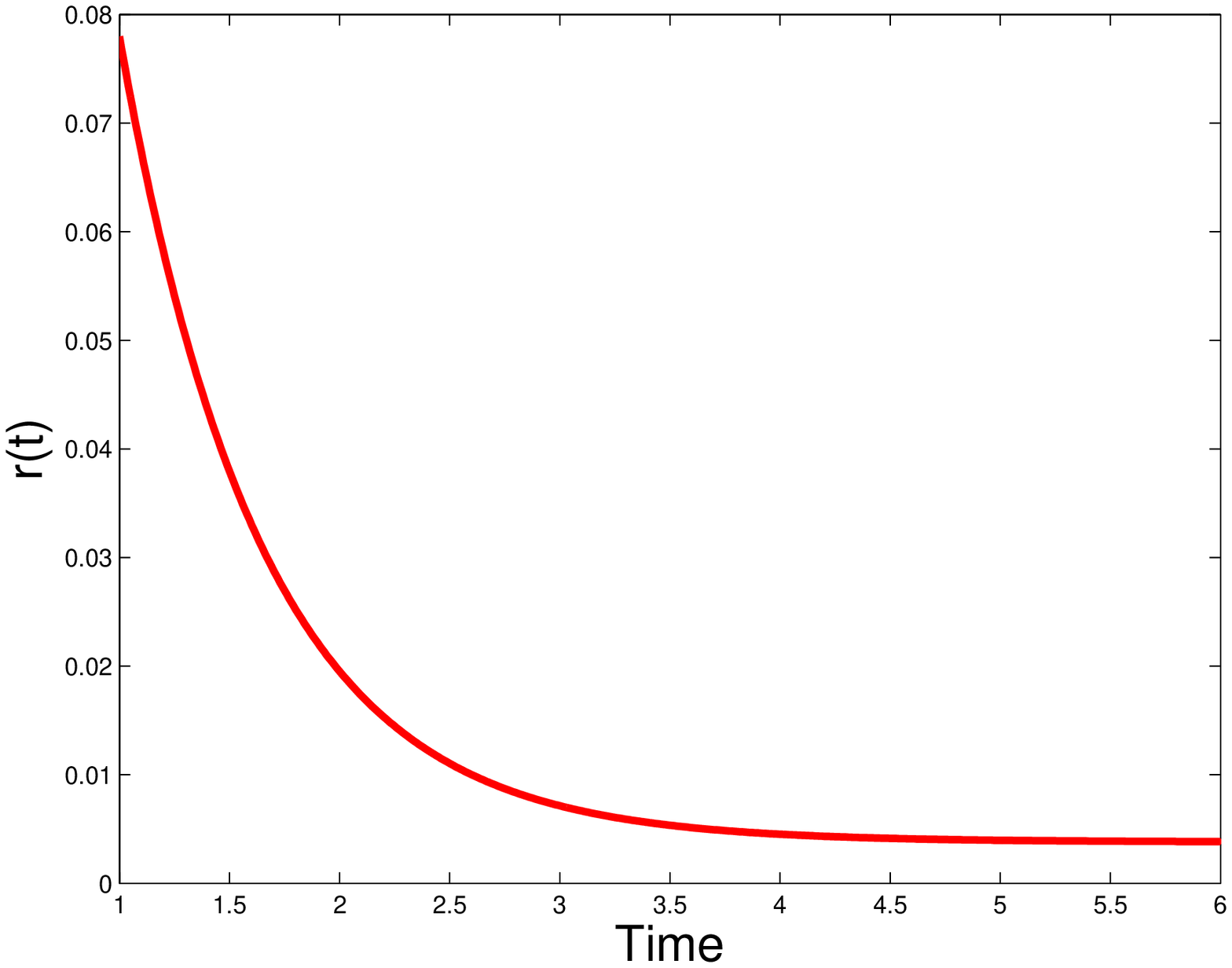}}
    \subfigure[$h(x)$ ]{
    \includegraphics[width=1.6in]{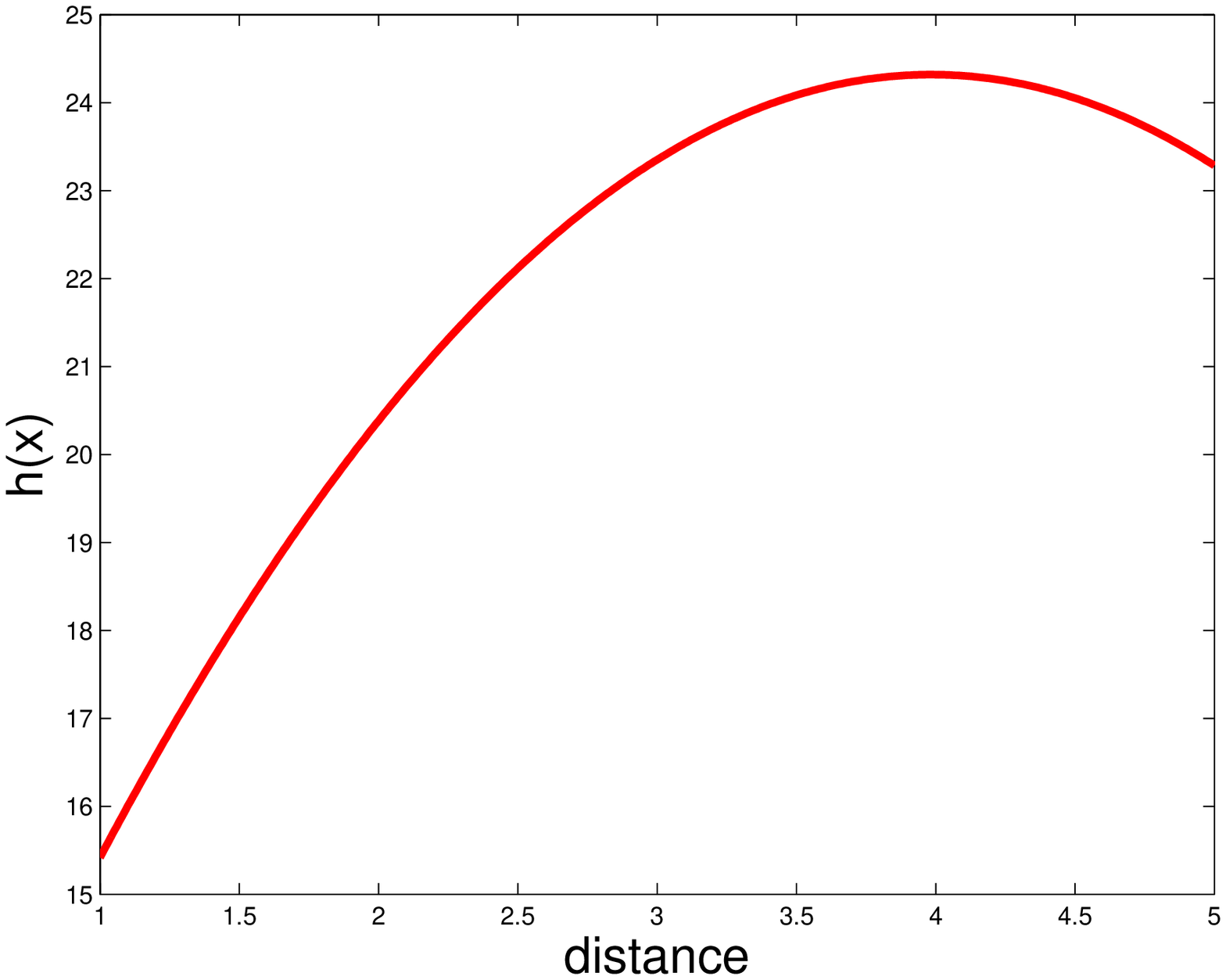}}
     \subfigure[Parameter values]{
\begin{tabular}{|c|c|} \hline

Parameter & value\\ \hline

d & 0.0020  \\ \hline

$\alpha$  & 1.5526 \\ \hline

$\beta$ & 0.0059  \\ \hline

$\gamma$ & 0.0780  \\ \hline

$\rho$ & -0.9478\\ \hline

$\sigma$ & 8.9149 \\ \hline

\end{tabular}
}
  \subfigure[Model Accuracy]{
\begin{tabular}{|c|c|} \hline

Distance & Average\\ \hline

1 & 97.88\%  \\ \hline

2 & 97.27\% \\ \hline

3 & 97.44\%  \\ \hline

4 & 96.20\%  \\ \hline

5 & 98.25\%  \\ \hline

Overall  &  97.41\% \\ \hline

\end{tabular}
}
 \caption{Accuracy of (\ref{eq2}) for the most popular news story in the Digg data set}
 \label{f332346.match.hop}
\end{figure}
We evaluate the performance of the proposed linear diffusive model by comparing the density calculated by the model with the actual observations in the Digg data set.  We numerically solved the model with Matlab. Figure~\ref{f332346.match.hop}[a] illustrates the performance of the linear diffusive model for the most popular story, $s1$.  Figure~\ref{f332346.match.hop}[b] gives the shape of $r(t)$ and Figure~\ref{f332346.match.hop}[c] gives the shape of $h(x)$. $h(x)$ is a concave down function with peak between $3$ and $4$, which is related to the neighbor distribution illustrated in Figure~\ref{fneighbor}. The shape of $h(x)$ suggests that there may exist highly influential users or opinion leaders at distance $4$ from the submitter of news $s1$. The corresponding parameters are listed in Figure~\ref{f332346.match.hop}[d]. The parameters are adjusted manually to best fit the actual data. The diffusion constant $d$ is relatively small because $d$ is the average diffusion rate for all distances. It also suggests that the structure-based process has a dominating impact on the information diffusion process. $\alpha$, $\beta$, $\gamma$ determine the shape of $r(t)$; and $\rho$ and $\sigma$ determine the peak of $h(x)$. The average accuracy at different distances are calculated for time $t=2,...,6$, and are provided in Figure~\ref{f332346.match.hop}[e]. The model can achieve high accuracy across distances.

 \cite{WWX2013} also studies the accuracy  of the model for describing all news stories in the Digg data set and examines whether the model can capture the heterogeneity features in information diffusion over the Digg network, we explore the overall accuracy of the linear diffusive model for all $133$ news stories with over $3000$ votes in the Digg data set.  Our results in \cite{WWX2013} illustrate that about $13\%$ of news stories can be described with accuracy higher than $90\%$. In total, about $60\%$ of news stories can be described with accuracy higher than $80\%$. The simulation is performed with a MATLAB auto fitting program. If we manually adjust parameters for each individual news story, higher accuracy can be achieved. For example, for the most popular news story, with manually adjusted parameters, the average accuracy can reach $97.41\%$, while with the automated parameter selection, the average accuracy is still greater than $90\%$. The high accuracy across all news stories with over $3000$ votes show strong evidence that the linear diffusive model captures the heterogeneity diffusion patterns of news and can be used as an effective approach to describe the news spreading in Digg.

\subsection{Logistic Model with Variable Content-based Diffusion}
In previous two subsections, we assume that the diffusion coefficient $d$ in the flux formula $$
J= -d \frac{ \partial I}{\partial x}
$$
is a constant. In fact, because of spatial heterogeneity of online users in social media, $d$ may be dependent on the distance $x$ from the source. In general,  $d$ may be a decreasing function of $x$ since interactions between different groups $U_x$ decrease dramatically as $x$ increases. Therefore, we use an exponential function $$
d=de^{-bx}
$$
to model the effect of spatial heterogeneity of online users in the content-based process in Fig. \ref{fprocess}.  Thus the following model combines the previous two models with variable contend-based diffusion.
\begin{equation}\label{eq3}
\begin{split}
&\frac{\partial I}{\partial t}= \frac{\partial(d e^{-b x} I_x)}{\partial x}+r(t)I(h(x)-\frac{I}{K})\\
& I(x, 1) = \phi(x), \;\;l< x < L  \\
&\frac{ \partial I}{\partial x}(l,t)=\frac{ \partial I}{\partial x}(L,t)=0, \;\;  t > 1 \\
&r(t)=A+Be^{-Ct}\\
\end{split}
\end{equation}
where
\begin{itemize}
\item $d$ represent the popularity of information; $b$ represents the decay of the popularity of information with respect to the friendship structure in social networks;
\item $K$ represents the carrying capacity, which is the maximum possible density of influenced users at a given distance;
\item $h(x)$ represents the heterogeneity of growth rate in distance $x$.
\end{itemize}

\begin{figure}[ht]
  \centering

    \includegraphics[width=3in]{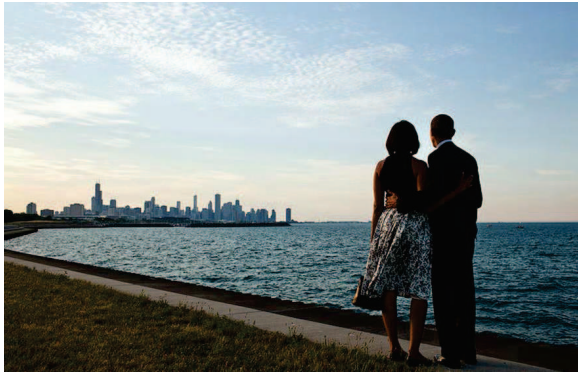}

 \caption{President Barack Obama tweeted the photo in Twitter at 10:20AM, Dec. 13, 2012. We choose the tweet as an example to validate (\ref{eq3}).}
  \label{obamaPhoto}
\end{figure}

\begin{figure}[ht]
  \centering
  \subfigure[Predicted (blue, solid) vs. Actual data (red, dotted)]{
    \includegraphics[width=3in, trim=1cm 5cm 1cm 5cm, clip=true]{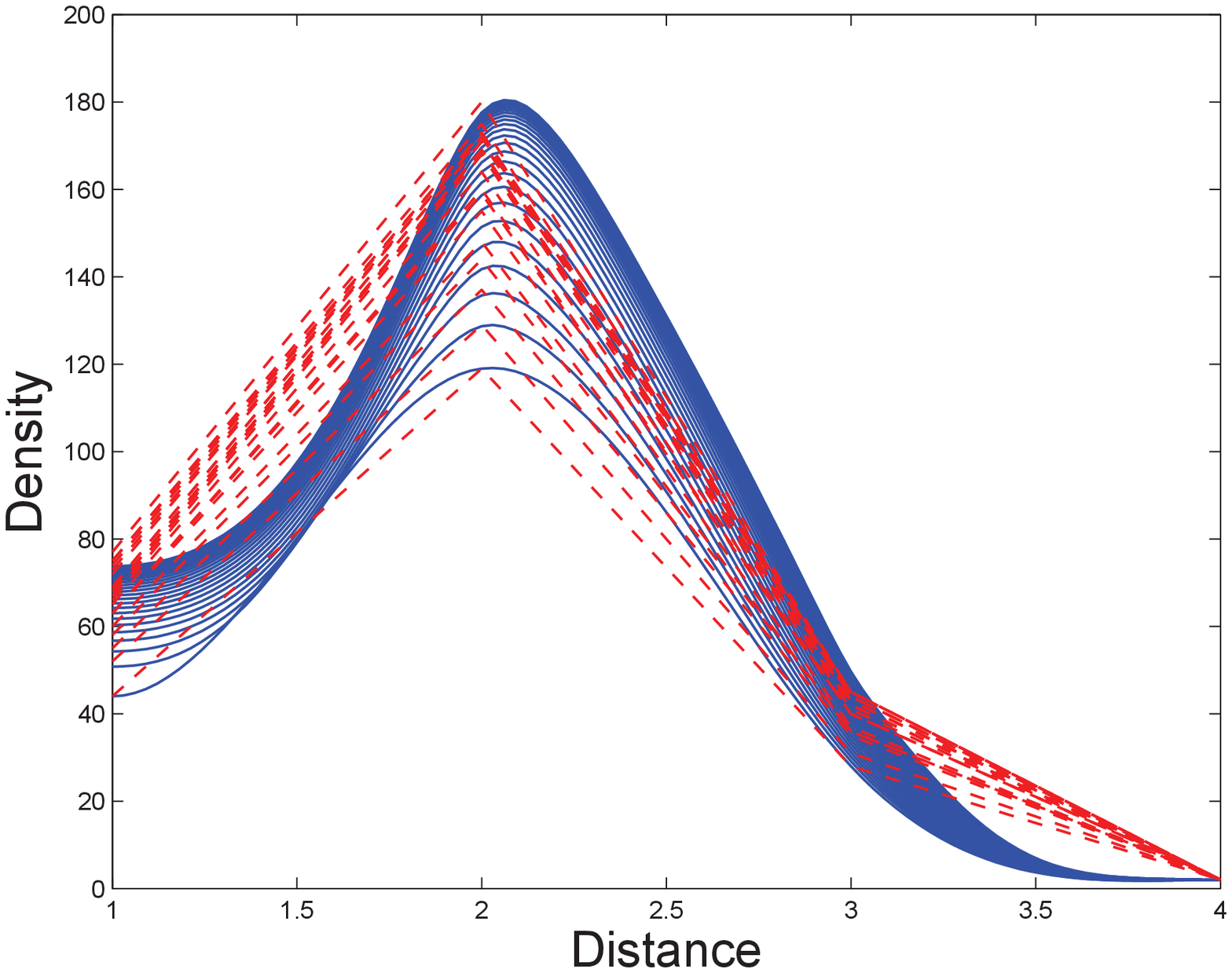}}
  \subfigure[$h(x)$ ]{
    \includegraphics[width=3in,trim=1cm 5cm 1cm 5cm, clip=true]{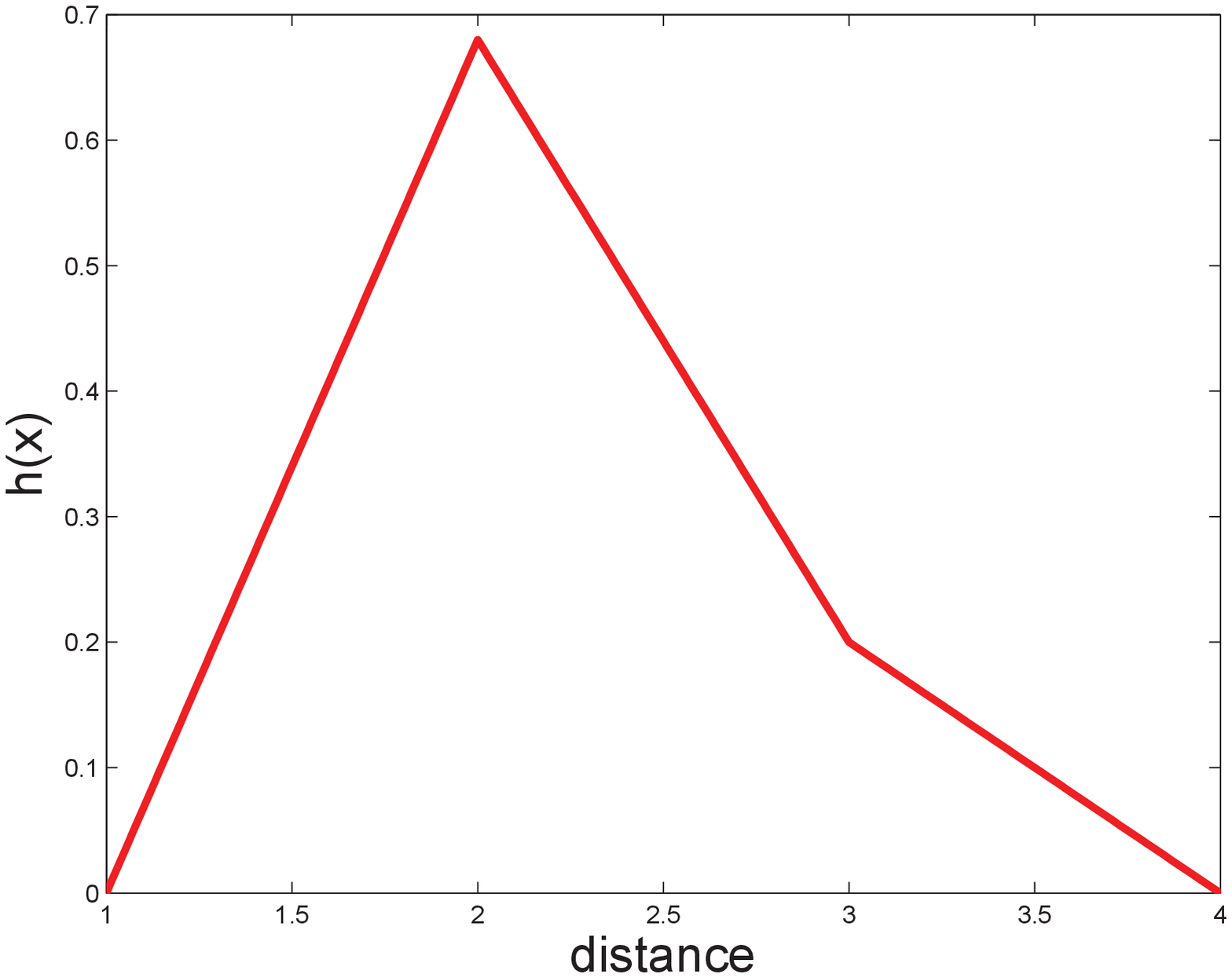}}
  \caption{Accuracy of (\ref{eq3}) for photo tweeted  by President Obama at Twitter }
 \label{obama.match.hop}
\end{figure}

\begin{figure}[ht]
  \centering

\begin{tabular}{|c|c|} \hline

Distance & Average\\ \hline

1 & 98.39\%  \\ \hline

2 & 98.75\% \\ \hline

3 & 94.11\%  \\ \hline

4 & 99.31\%  \\ \hline
Overall  &  97.64\% \\ \hline

\end{tabular}

 \caption{Accuracy of (\ref{eq3}) at $x=1,2,3,4$ for photo tweeted by President Obama}
 \label{table.obama.match.hop}
\end{figure}

\subsubsection{Accuracy of Logistic Model with Variable Content-based Diffusion}
We evaluate the performance of the logistic model with variable diffusion by comparing the density calculated by the model with the actual observations in the Twitter dataset.  We choose the photo in Figure~ \ref{obamaPhoto} tweeted by President Barack Obama on December 13, 2012 as an example to verify accuracy of the model. Based on the Twitter user list and social network graphs collected in October 2009, we find 383 Twitter users have retweeted this photo. Among these 383 users, 96 users has a distance of 1 to the source of the tweet message, 226, 58, 2, and 1 users have a distance of 2, 3, 4, and 5 respectively.  We compare the number of users retweeting the photo at the different distances and a time frame from $t=1$ to $t=15$.

We numerically solved the model with Matlab. Figure~\ref{obama.match.hop} (a) illustrates the predicting results for the photo tweeted by President Barack Obama
with the proposed model for the time frame from $t=1$  to $t=15$ hours, where the $x$-axis is the distance measured by {\em friendship hops}, while
the $y$-axis represents the number of retweeted users within each distance. The red-lines represent the actual number of people retweeting the photo at various time increments. The blue curves represent the model used to predict the information diffusion based on the PDE model. The red-dotted
lines denote the {\em actual} observations for the number of retweeted users. The mathematical model represented the diffusion of President Obama's tweet reaches an overall accuracy of 97.64 \%, shown in the Figure ~\ref{table.obama.match.hop}.   These results were obtained with $d=1, b=3, K=300$, $r(t) = 0.3 + e^{-2t}$, $h(x)$ function shown in Figure~\ref{obama.match.hop} (b). Figure~\ref{obama.match.hop} (b) illustrates a peak at $x=2$, which can indicate that President Obama's tweeted the photo is most popular within the Twitter users at distance two.

Therefore, the spatial models (\ref{eq1}), (\ref{eq2}) and (\ref{eq3}) can achieve high accuracy. While  (\ref{eq2}) is a linear model and captures the behavior of news spread within a few hours, nonlinear models (\ref{eq1}) and (\ref{eq3}) can predicate news spread for a longer time frame.  $h(x)$ in (\ref{eq2}) and (\ref{eq3}) reflects the spatial heterogeneity of online users with respect to distance $x$.  In (\ref{eq1}) it is assumed that $h(x)$ is constant. (\ref{eq3}) takes into consideration  of the fact that the diffusion coefficient $d$ is a decreasing function of $x$, which is a constant in both (\ref{eq1}) and (\ref{eq2}).  From the experiments above, all three models can achieve extremely high accuracy.

\subsection{PDE model with Distance Based on Shared Interests}\label{interest}

While we have used the number of friendship hops as a natural distance,  an alternative approach could be
{\em interest distance}, for measuring the distance
between two users through their shared interests on information or content
in social networks. Given two social networks users $a$ and $b$,
we use $C_{a}$ and $C_{b}$ to denote the set of contents the
users $a$ and $b$ have interacted with, respectively. In the
context of {\tt Digg} social networks, $C_{a}$ represents
all the news stories the user $a$ have voted or digged.
The {\em shared interest} $d_{a,b}$ between users $a$ and $b$ is
defined as:
\begin{equation}
d_{a,b} = 1 - \frac{C_a \cap C_b}{C_a \cup C_b}
\end{equation}
where $C_a \cup C_b$ is the number of the total contents that
either user $a$ or user $b$ has interacted with and $C_a \cap C_b$
is the number of the shared contents that both users $a$ and $b$
have interacted with. Essentially the interest distance
quantifies the degree of the shared interests among two
users. An information originating from the source $s$ is likely
to influence users who have small interest distances
to the source due to the shared interests.  In a recent work \cite{WWX2012inter} we introduced an effective algorithm to identify the shared interests in online social networks. With the distance, all previous models can be modified to reflect how information flows from these who share more common interests to those with less common interests. We shall discuss more about the problem in a future work.

\section{Modeling Complex Interactions}\label{interaction}

Information diffusion process over online social networks could be influenced
by more complex interactions between users and information, e.g., information originating
from multiple sources and competing information from different political campaigns. Myers et al. \cite{MyersCooper2012} (also see Guille \cite{GuilleSurvey2013}) studies cooperation and competition in information diffusion in the Twitter network using a statistical model.  In this section, we start with a number of simple models to illustrate how to incorporate complex interactions in the spatial-temporal setting. We are in the process to refine and validate these models with real datasets.


\subsection{Multiple Information Sources}
In an online social network, same information could originate from multiple sources.
Often, breaking news stories, emergency events, and controversial topics are initiated by a number of different news sources; Multiple users tweeting the final result for a sport game in Twitter. News from multiple sources often increase its spreading speed and coverage. It is more practical to understand the diffusion patterns of multi-source information in social media.

In a recent paper \cite{PXWW2013} by Peng and the authors, we use the linear diffusion model (\ref{eq2}) to predict the information diffusion process of multi-source news spread in Digg. \cite{PXWW2013} studies
the basic characteristics of the diffusion process of multi-source information.  The distance metric in \cite{PXWW2013} is intuitively defined as the minimum shortest path between a user and multiple sources. The distance definition reflects the fact that while a user of social media could be influenced by a set of news sources via different multiple
paths, but the nearest source to which the user has the minimum
friendship hop has the highest probability of influencing the
behavior of the given user due to the smallest number of
friendship hops. The predication accuracy of the linear diffusion model (\ref{eq2}) for multiple news sources is validated  with the same data-set from Digg. Our experiments in \cite{PXWW2013} show that the model can describe the
most representative news stories initiated from multiple sources
with an accuracy higher than 90\%, and can achieve an average
accuracy around 75\% across all multi-source news stories in
the data-set. These results confirm that our approach with reaction-diffusion equation is able to describe
and predict the spreading patterns of multi-source information with high accuracy.

The effect of multiple sources can be also viewed as mutualism in biological systems where two or more organisms of different species biologically interact in a relationship in which each individual derives a fitness benefit (i.e., increased or improved reproductive output). Such interactions in biological systems are known as cooperation systems ~\cite{Murray1989}.  The number of influenced users at any location and time can dependent on multiple distances and time. Therefore dynamical equations can predicate how the information spreads in social media with interactions between multiple directions and channels.  If we consider two news sources, then the two pieces of news spreads with logistic growth independently along with additional effect from another. Let $u_i$ be the density of the information from different sources. The positive effects can be modeled by terms $\alpha_1 u_1 u_2$ and $\alpha_2 u_1 u_2$. As a result, the following simple model can be a starting point to address the impact between multiple sources:

\begin{equation}\label{cooperative}
\begin{split}
&\frac{\partial u_1}{\partial t}=d_1 \frac{\partial^2 u_1}{\partial x^2}+r_1(t) u_1(1-\frac{u_1}{K_1}) +\alpha_1 u_1 u_2\\
&\frac{\partial u_2}{\partial t}=d_2 \frac{\partial^2 u_2}{\partial x^2}+r_2(t) u_2(1-\frac{u_2}{K_2}) +\alpha_2 u_1 u_2,
\end{split}
\end{equation}
where

\begin{itemize}

\item $d_1, d_2$ represent the popularity of the two pieces of information.
\item $r_i(t), i=1,2$ represents the intrinsic growth rate of influenced users with the same distance, and measures how fast the information spreads within the user groups with the same distance;
\item $K_i, i=1,2$ represents the carrying capacity, which is the maximum possible density of influenced users;
\item $\alpha_1$ measures the positive effect of news $u_2$ on $u_1$ and $\alpha_2$ measures the positive effect of news $u_1$ on $u_2$.

\end{itemize}
In addition to study solutions of (\ref{cooperative}) along with appropriate boundary and initial conditions, we are also interested in how fast the information spreads in online social networks with multiple sources. We will discuss it in Section \ref{multipleSource} where we assume that the underlying domain is from $-\infty$ to $\infty. $

\subsection{Competing Information }
It is common to observe completing information from different political
or product campaigns, where information needs to compete with each other
to maximize its own influences on people through online social networks.
To characterize the density of users that are influenced by a certain
competition, we may study competing news by the Lotka-Volterra competition models ~\cite{Murray1989}. Competition models
have been extensively studied in mathematical biology ~\cite{Murray1989} where interactions between organisms or species can lower the presence of another due to limited resources (such as food, water, and territory).  A simple mathematical model to describe news competition may be the same as equation (\ref{cooperative}) except that
the effect of the competition
is negatively proportional to the number of influenced users.
\begin{equation}\label{CompetingInfo}
\begin{split}
&\frac{\partial u_1}{\partial t}=d_1 \frac{\partial^2 u_1}{\partial x^2}+r_1(t) u_1(1-\frac{u_1}{K_1}) -\alpha_1 u_1 u_2\\
&\frac{\partial u_2}{\partial t}=d_2 \frac{\partial^2 u_2}{\partial x^2}+r_2(t) u_2(1-\frac{u_2}{K_2}) -\alpha_2 u_1 u_2,
\end{split}
\end{equation}
Here $\alpha_1$ measures the competition effect of news $u_2$ on $u_1$ and $\alpha_2$ measures the competition effect of news $u_1$ on $u_2$. Thus the model may be able to capture the logistic growth of each information, and also quantify the effect of the competition on the diffusion process. Some relevant and practical questions,  such as which news can win the competition,  can be answered by studying the competition models in Section \ref{ComputingSource}.

\subsection{Spatial Epidemiological Models}
It is commonly accepted that dynamical models to describe disease spreading can be used to study information. There are a considerable body of research work modeling information diffusion based on techniques for modeling infectious diseases ~\cite{Newman2003}. In general, epidemiological models are neither cooperative nor competitive and more challenging to study. In principle, the nodes of a network are classified into several classes (i.e. states) and focus on the evolution of the proportions of nodes in each class. $SI$ and $SIR$ are the two basic models, where $S$ stands for "susceptible", $I$ for "infected" (i.e. adopted the information) and $R$ for recovered (i.e. refractory). In both cases, nodes in the $S$ class switch to the $I$ class due to influence of their neighbor nodes. Then, in the case of $SI$, nodes in the $I$ class switch to the $S$ class, whereas in the case of $SIR$ they permanently switch to the $R$ class. The percentage of nodes in each class is expressed by simple differential equations. Both models assume that every node has the same probability to be connected to another and thus connections inside the population are made at random.

However, most of the work on social media has largely concentrated on collective analysis and involves only ordinary differential equations. With the new metric concept between users we introduced our recent papers, spatial effects can be incorporated and partial differential equations come into play.  In particular, spatial models take into consideration of the external influence.  Many similar concepts and models in epidemiology can be further modified and expanded to study information diffusion in online social networks.  For social media, $S$ represents the density of susceptible users  at time $t$ and distance $x$ in $U_x$ and $I$ represents the density of influenced users at time $t$ and distance $x$ in $U_x$.  The following $SI$ model is a simple example how spatial infectious disease model can be used to study online social networks.
\begin{equation}\label{SIS}
\begin{split}
&\frac{\partial S}{\partial t}=d_1 \frac{\partial^2 S}{\partial x^2}-r(t) \frac{SI}{S+I} \\
&\frac{\partial I}{\partial t}=d_2 \frac{\partial^2 I}{\partial x^2}+r(t) \frac{SI}{S+I}
\end{split}
\end{equation}
where $r(t)$ is the rate of influence. $S+I$ appears in the denominator because it is not necessarily constant in spatial models.  One important concepts to describe the interactions between user groups is the rate of influence $r(t)$ which is similar to the force of infection in epidemiology.  The choice of the rate of influence is largely dependent on news and user classifications. Data mining techniques and graphical model can significantly improve the selection of the parameters. We shall determine the parameters for news from Twitter in a future work.

\subsection{Multiple Communication Channels}
Most of the research on information diffusion is limited to study information diffusion from isolated single media sites or factors, such as friendship hop etc. However, many online social networks involve multiple connections between any pair of nodes (such as friendship, family relationship, geographic location and other demographic characteristics). In addition, news propagation in one social media site often affects that of another social media site, for example, Facebook vs. Twitter. A theoretical framework for understanding of information flow in online social networks over multiple communication channels is still missing. With the framework to model information diffusion in social media with PDE models in \cite{WWX2012,WWX2013}, it is possible to use partial differential equations defined on multidimensional domains to analyze, characterize and predict spatial-temporal dynamics of information diffusion with multiple communication channels.  The governing partial differential differentials are systems of reaction-diffusion equations defined on multidimensional domains. For example, with almost the same setting as in (\ref{eq1}),  a simple diffusive logistic equation defined on multidimensional domain $\Omega$ may be able to describe information diffusion over social media with multiple communication channels

\begin{equation}\label{eqmultiple}
\begin{split}
&\frac{\partial I}{\partial t}=d \Delta I+r I(1-\frac{I}{K})\\
& I(x, 1) = \phi(x), \;\; x \in \Omega  \\
&\frac{ \partial I}{\partial n}=0, \;\;  \text{ on } \Omega \times (1, \infty) \\
\end{split}
\end{equation}
here $\Delta$ is the Laplacian operator defined on multidimensional domain $\Omega \subset R^{n}$. The shape of $\Omega$ can be determined by the correlation of the communication channels. For example, if we are interested in studying a specific region with two unrelated related channels, a rectangle region may be plausible to describe the combined effect of the two channels on the spreading of news in a social media site. Systems of equations such as (\ref{cooperative}) or (\ref{SIS}) can also be defined on multidimensional domains to model information diffusion with multiple communication channels.  We shall collect data from multiple social media sites  to validate the model. The results can be used to help uncover impacts of multiple communication channels and multiple social media sites on information diffusion in online social networks.

\section{Model Analysis and Discussion}\label{mathh}

We have presented a number of partial differential equation models to characterize spatial-temporal patterns in social media. The spatial-temporal models are reaction-diffusion equation models built on intuitive cyber-distance in online social networks. The basic mathematical properties such as the existence, uniqueness and positivity of the solution of the models can be established from the standard theorems for parabolic PDEs \cite{cc,halsmith}.  Many other mathematical properties remain to be investigated. Not only mathematical analysis of the models can validate the models, but also reveal insights of information flow over social media through mathematical properties of the solutions of PDE models.  In this section, instead of giving technical details, we briefly highlight a number of interesting mathematical problems and their social implications.

\subsection{Information Classification and Parameter Selection}
The PDE models we discussed here are general principles governing information diffusion over social media. The type of information plays a significant role in the diffusion process of information over social media. Online users more likely spread news they are interested in.   Hence, it is necessary to differentiate and classify types of information. As the first step, we select the best parameters to maximize the accuracy of the PDE models in comparison with real datasets. The chosen parameters can be used to classify the information. Once we are able to classify or categorize information, the PDE models can be used to predicate how information spread through solving the PDE models along with appropriate parameters according to the category of information. Graphical models and data mining techniques will be utilized to categorize the information that spreads over online social networks.  Mathematical studies such as stability and bifurcation analysis can determine parameter ranges or cut-off points for information spreading patterns to preserve predicable trajectories or undergo sudden changes.  As we can see from the next section, one of the important mathematical concept, positive principal eigenvalue of PDE models, plays a pivotal role in the determination of stability and bifurcation of the model below.  As such, mathematical analysis of the PDE models will solidify the foundation for information classification and parameter selection.

\subsection{Stability and Bifurcation }
Some information can spread in social media for a long period time even years \cite{Cha2009}.  The simulations with real data collected from social media suggest that our models can achieve high accuracy over an extended time frame, such as  more than 40 hours. We are interested in how the size and structure of the underlying network influence asymptotic behavior of news spreading, coexistence of multiple competing news, or decay rate of news over an extended time frame.  Stability analysis of the models can predict trajectories of information flow over social media under small perturbations of initial conditions. Bifurcation analysis of the models can reveal how changes of parameter values of the models may cause qualitative or topological changes in the behavior of the models.

Long term behavior of the solutions of the model may depend on their corresponding elliptic equations and parameters of reaction-diffusion equations, reflecting the popularity or spreadiblity of news in social media.  In the context of spatial ecology and physics, extensive research on eigenvalue analysis, stability, effects of boundary conditions, evolution of dispersal has been done. \cite{cc,lou} and references therein include some recent developments in spatial ecology. In a recent paper \cite{DaiMaWang2013}, Dai, Ma and the first author study stability and bifurcation of the following model arising from online social networks
\begin{equation}\label{MDrobinBC2}
\left\{
\begin{array}{l}
u_t-(a(x)u_{x})_x=\lambda r(t) u \left(h(x)-\frac{u}{K}\right), ~~~~~\,\,\,\,\, t>1,\,\, l<x<L,\\
u(1,x)=u_0(x),~~~~~~~~~~~~~~~~~~~~~~~~~\,\,\,\,\,\, l\leq x\leq L,\\
u_x(t,l)=0,\,\,u_x(t,L)+\alpha u(t,L) =0,\,\,\,t>1,
\end{array}
\right.
\end{equation}
where $a(x)=de^{-bx}$, and $\alpha, d, b $ are positive constants, $h(x)$ is positive and $r(t)$ is a decay function approaching $r_{\infty}$ as $t \to \infty.$  The parameter $\lambda$ can be interpreted as a scale or factor of $r(t)$.  The Robin boundary condition at $x=L$ reflects the fact that there is an exchange of information at the boundary. For $\alpha >0$, it indicates the flux $-u_x(t,L)$ is positive and therefore information flows to the right. The simulations with real data set from Digg suggests that the Robin boundary condition can even achieve high accuracy.

The steady state solution equation of (\ref{MDrobinBC2}) satisfies
\begin{equation}\label{DL4}
\left\{
\begin{array}{l}
-(a(x)u')'= \lambda  r_{\infty} u \left(h(x)-\frac{u}{K}\right), \,\, l<x<L,\\
u'(l)=0,\,\,u'(L)+\alpha u(L)=0,
\end{array}
\right.
\end{equation}
\cite{DaiMaWang2013} studies the global bifurcation and stability of (\ref{MDrobinBC2}) and its implication to social media.  Consider the following associated eigenvalue problem with the parameter $\mu$
\begin{equation}\label{DL5}
\left\{
\begin{array}{l}
-(a(x)u')'=\mu  h(x) u, \,\, l<x<L,\\
u'(l)=0,\,\,u'(L)+\alpha u(L)=0,
\end{array}
\right.
\end{equation}
It is known that (\ref{DL5} ) has the positive eigenfunction and principal eigenvalue $\mu_1^+$, which is determined by
\begin{equation}
\frac{1}{\mu_1^+}=\max_{u\in W^{1,2}([l,L]), u\neq 0}\left[\frac{\int_l^L h(x)u^2\,dx}{\int_l^L\left\vert a u'\right\vert^2\,dx+\alpha u^2(L)}\right].\nonumber
\end{equation}
It is shown in ~\cite{DaiMaWang2013} that (\ref{DL4}) has an unbounded branch of positive solutions bifurcating from ($\frac{\mu_1^+}{r_{\infty}},0$).
The bifurcation result for (\ref{MDrobinBC2}) can be derived from well known general bifurcation results (see e.g. \cite{cc}). It is also shown  in \cite{DaiMaWang2013}  that if $\lambda> \frac{\mu_1^+}{r_{\infty}}$, then (\ref{MDrobinBC2}) has a positive steady solution which attracts all its solutions as $t$ goes to infinity; while $\lambda< \frac{\mu_1^+}{r_{\infty}}$ implies all its nonnegative solutions go to zero.

Information diffusion, in particular, news diffusion over social media is often time sensitive. Reaction-diffusion equations arising from online social networks involves a decaying function $r(t)$. Most of the existing research focus on the cases that $r(t)$ is constant \cite{cc} or periodic \cite{hess}. Some researchers also study eigenvalue, stability and persistence of nonautonomous parabolic PDEs \cite{Mierczynski2000,LANGA2009}.  Mathematical analysis of associated eigenvalue and bifurcation problems can help identifying thresholds for the change of social dynamics.  Mathematically, eigenvalues, stability, bifurcation and persistence of the reaction-diffusion equations are obtained for $h(x)$ is positive and there are some interesting results and more challenging problems when $h(x)$ may take negative value \cite{cc,lou}.  In the context of social media, $h(x)$ may be negative for some $x$ in particular when negative news or spam are involved as many online users will delete them.   With the availability of real data from social media, we are in a position to study the challenging problems from both theoretical and practical aspects, identify conditions for stability and persistence, and equally importantly, verify the conditions through the real data sets collected from social media.

The study of spatial heterogeneity on information diffusion in social media has significant theoretical and practical implications. For example, since $h(x)$ represents the adoption rate of information for the group users whose distance away from the origin is $x$, the shape of $h(x)$ may contribute to locate the so-called the most influenced users or opinion leaders in social media. Other related interesting problems include maximizing the total influenced users for certain classes of $h(x)$. The issue are of interest as it has commercial potentials and social implications. Numerous research on this issue has emerged in recent years to design efficient algorithms for detecting opinions from corpus of data \cite{GuilleSurvey2013}. Our PDE models provide a new framework to design detection algorithms by studying mathematical properties of $h(x)$. As a result, recent theoretical developments on non-linear partial differential equations can facilitate the research and development of the important social problem.

\subsection{Free Boundary Value Problems}\label{freebvp}

In our previous models (\ref{eq1}),(\ref{eq2}), (\ref{eq3}) no flux boundary condition is assumed with the understanding of no formation across the boundaries at $x=l,L$. This might be reasonable when the number of the groups, or the maximum of $x$ are small. If the number of the groups are large, the boundary of influenced online users may change as news spreads. To describe the change of boundary with respect to time $t$, recently, Lei, Lin and the first author \cite{LZW1013} proposed and studied the following free boundary model to describe the spreading of news in online social networks
\begin{eqnarray}
\left\{
\begin{array}{lll}
u_{t}-d u_{xx}=r(t)u(1-\frac u K),\; & t>0, \ 0<x<h(t),  \\
u_x(t,0)=0,\; u(t, h(t))=0,\quad & t>0, \\
h'(t)=-\mu u_x(t,h(t)),\quad &t>0,\\
h(0)=h_0,\,  u(0, x)=u_{0}(x),\; &0\leq x\leq h_0,
\end{array} \right.
\label{f3}
\end{eqnarray}
where  the initial function $u_{0}(x)$ satisfies \begin{equation}
u_0\in \Sigma(h_0)=\{\phi \in  C^{2}([0, h_0]): \quad \phi'(0)=\phi(h_0)=0,\ \ \textrm{and} \ \phi>0\ \textrm{in}\ [0, h_0)\}.
\label{a11}
\end{equation}
$x=h(t)$ is the moving boundary to be determined and represents the spreading front of news (such as movie recommendation) among users.  $h'(t)=-\mu u_x(t,h(t))$ is the Stefan condition, where $\mu$ represents the diffusion ability of the information in the new area. Let $r_{\infty}=\lim_{t \to \infty}r(t)>0$. It is well known that the Stefan conditions have been used in many areas when phase transitions in matters such as ice passing to water and other biological problems.

It was shown in \cite{LZW1013} that the free boundary $x=h(t)$ is increasing.  Further, it was shown that the information traveling either lasts forever or suspends in finite time.
In addition, the impact of the initial condition of news on its spread over online social networks is discussed. Let $u_0=\lambda\varphi$ for some $\varphi$ belongs to $\Sigma(h_0)$, it was shown in \cite{LZW1013} that if $\lambda$ is sufficiently small, the information vanishing must occur. Then it was shown that there exists a threshold $\lambda^*$ which is dependent on $\varphi\in\Sigma(h_0)$ such that when $\lambda>\lambda^*$, the information with the initial data $u_0=\lambda\varphi$ travels in the whole distance. Otherwise, the information vanishing happens.

Finally, if the information spreading happens, the expanding news front $x=h(t)$ moves at a constant speed $k_0$ for large time.  It is shown in \cite{LZW1013} that the following relation holds
\begin{equation}\label{speeds}
  \lim_{\frac{\mu K}{d}\to \infty} \frac{k_0}{\sqrt{r_{\infty}d}}=2.
\end{equation}
(\ref{speeds}) indicates that the asymptotic traveling speed $k_0$ is close to $2 \sqrt{r_{\infty}d}$, which is also called the minimum speed of (\ref{f3}) for the Fisher's equation  as we shall discuss in depth in Section \ref{travl}. The asymptotic traveling speeds of news fronts from free boundary problems and the minimum speeds from traveling wave solutions in Section \ref{travl} can provide a theoretical guide for how to maximize or control information propagation in online social networks. Several free boundary value problems related to (\ref{eq3}) remain to be mathematically studied. For example, information diffusion with multiple channels can give rise to partial differential equation (\ref{eqmultiple}) defined in more complex domains. In such a setting, it is interesting to investigate how news front $h(t)$ changes at different directions.

\subsection{Traveling Wave Solutions and Spreading Speeds }\label{travl}

In general, news diffusion in social media is time-sensitive and the influence of news decays drastically as time elapses. However, some information can take a longer period of time to spread in social media.  Cha et al. \cite{Cha2009} examined the aggregate growth patterns of two sets of 5,346 and 897 photos in the Flickr social network (http://www.flickr.com/) that are older than $1$ year and $2$ years respectively.  It is found in \cite{Cha2009} that the long-term trend in the number of cumulative fans exhibits a pattern of steady linear growth.  Many photos show quick rise in popularity during the first few days after being loaded. However,  most of the pictures exhibit a period of steady linear growth after the first few $(10-20)$ days. More importantly the linear-growth is sustained over extended periods of time. The growth rate continues to increase even after $1$ or $2$ years. Thus, for long-term propagation of information, we may choose distance metrics in a way that online users can be embedded in the whole $x$-axis and the source of information can be viewed from either from $-\infty$ or $\infty$. Further, the parameters may be chosen to be independent of time $t$.  As such, it is meaningful to discuss long-time behavior and traveling solutions of the reaction-diffusion systems for information diffusion in online social networks.  A traveling wave solution often represents a transition process connecting two steady states of interactive populations.  Traveling wave fronts of partial differential equations are solutions of the form $u(x+ct)$ that has a fixed shape and translate at a constant speed $c$ as time evolves. The wave speed $c$ are interpreted as the rate of spread of the introduced population in biology. The theoretical results on traveling wave solution of reaction-diffusion equations has successfully predicted spread rates of some introduced species.

For the long-time behavior and spatial spread of an advantageous gene in a population, Fisher \cite{Fisher} and  Kolmogorov, Petrowski, and Piscounov  \cite{Kolmogorov} studied the nonlinear parabolic equation
\begin{equation}\label{eq001}
u_t=du_{xx}+f(u)
\end{equation}
here, $u(x, t)$ represents the population density at location
$x$ and time $t$ and $f(0)=f(1)=0$ and $f(u)>0$ with no Allen effort.  Traveling wave fronts of (\ref{eq001}) are of interest since they enable us to better understand how a population propagates.  It was shown that (\ref{eq001}) has a traveling wave solution of the form $u(x+ct)$ if and only if $|c| \geq c^*$  and the minimum speed of propagation for (\ref{eq001}) is $c^*$
where  $$c^*=2\sqrt{f'(0)d}$$
This basic formula $c^*=2\sqrt{f'(0)d}$ establishes the speeding spreads for nonlinear parabolic equations and indicates the rate of spread is a linear function of time and  that it can be predicted quantitatively as a function of measurable life history parameters.

In spatial biology and epidemiology, it is of great interest to estimate how fast a species or infectious disease spread within a population. Building on the mathematical foundation for the theory of spreading speeds for cooperative systems by Weinberger et al. \cite{Weinberger2002-1},  the first author \cite{Wang2010jns} discussed spreading speeds for a large class of systems of reaction-diffusion equations which are not necessarily cooperative through  analysis of traveling waves via the convergence of initial data to wave solutions. In particular, \cite{Wang2010jns} provides a practical approach to calculate the propagation speed based on the eigenvalues of the parameterized Jacobian matrix of its linearized system at the initial state. Here we follow the direct derivation in \cite{Wang2010jns} from the perspective of traveling wave solutions. Let us consider a system of reaction-diffusion equations with zero and another positive equilibria.
\begin{equation}\label{eq11}
\u_t=D\u_{xx}+\f(\u) \text{ for } x \in \mathbb{R},\; t\geq 0
\end{equation}
where $\u=(u_i)$, $D=\text{diag} (d_1, d_2, ...,d_N),  d_i>0 \text{ for } i=1,...,N$
$$\f(u)=(f_1(u),f_2(u),...,f_N(u)),$$
We are looking for a traveling wave solution $\u$ of (\ref{eq11}) of the form $\u=\u(x+ct), \u \in C(\mathbb{R}, \mathbb{R}^N)$ with a speed of $c$ . Substituting $\u(x,t)=\u(x+ct)$ into (\ref{eq11}) and letting $\xi=x+ct$, we obtain the wave equation
\begin{equation}\label{eq211}
D\u''(\xi)-c\u'(\xi)+\f(\u(\xi))=0\text{ for } \xi \in \mathbb{R}.
\end{equation}
Now if we look for a solution of the form
$(u_i)=\big ( e^{\lambda \xi}\eta^i_{\lambda}\big),\lambda>0, \eta_{\lambda}=(\eta^i_{\lambda})>>0$ for the linearization of (\ref{eq211}) at an initial equilibrium   at the origin,
we arrive at the following system
\begin{equation*}
\text{diag}(d_i \lambda^2 -c \lambda)\eta_{\lambda}+\f'(0)\eta_{\lambda}=0
\end{equation*}
which can be rewritten as the following eigenvalue problem
\begin{equation}\label{egenvalue}
\frac{1}{\lambda}A_{\lambda}\vect{\eta_{\lambda}}= c \vect{\eta_{\lambda}},
\end{equation}
where
\begin{equation*}
A_{\lambda}=(a^{i,j}_{\lambda})=\text{diag}(d_i\lambda^2)+\f'(0)
\end{equation*}
 Let $\Psi(A_{\lambda})$ be the spectral radius of $A_{\lambda}$ for $ \lambda \in [0, \infty)$,  $$
\Phi(\lambda)=\frac{1}{\lambda} \Psi(A_{\lambda})> 0.
$$
In \cite{Wang2010jns}, under assumption that $\f'(0)$ has nonnegative off diagonal elements and some other conditions, it was shown that $\Phi(\lambda)$ is a convex-like function and $\Phi(\lambda)$ goes to $\infty$ at both of $0$ and $\infty$. Therefore $\Phi(\lambda)$ assumes the minimum over the domain $(0, \infty)$, that is
\begin{equation}\label{mini}
c^*=\inf_{\lambda>0}\Phi(\lambda)>0
\end{equation}
The value of $c^*$ reflects information propagation speeds within a population.  $c^*$ is often called the minimum speed for systems that are linearly determinate. However, it is a challenging mathematical problem to prove a system is linearly determinate, in particular for non-cooperative system. It is known that cooperative systems and a few of other type of systems are linearly determinate \cite{Weinberger2002-1,Wang2010jns}. In the next three subsections we will discuss spreading speeds of systems for multiple sources, competing information and epidemiological process. We are in the process to valid the theoretical results with real data. Nevertheless, these results can serve a starting point to quantify information diffusion spreading in online social networks.

\subsubsection{Propagation Speeds for Multiple  Sources} \label{multipleSource}
As we discuss before, certain information such as photos in the Flickr social network can take a long period of time to spread and exhibits a pattern of steady linear growth. In this case, we can assume the coefficient in (\ref{cooperative}) are all positive constant. (\ref{cooperative}) is a cooperative system since its Jacobian $$
\left(
  \begin{array}{cc}
    r_1-2\frac{r_1u_1}{k_1}+ \alpha_1 u_2 & a_1u_1 \\
    \alpha_2 u_2 & r_2-2\frac{r_2u_2}{k_2}+ \alpha_2 u_1 \\
  \end{array}
\right)
$$
has nonnegative off-diagonal elements for $u_1, u_2 \geq 0$.  Now we can use (\ref{mini}) to calculate the minimum speed of information propagation in social media for a longer period of time where $r_i, \alpha_i$ are all positive constant. We are interested in a transition process connecting two equilibria $(0,0)$ and $(e_1, e_2)$ of (\ref{cooperative})  where $$e_1=\frac{k_1r_2(\alpha_1 k_2 +r_1)}{r_1r_2-\alpha_1\alpha_2 k_1k_2}, e_2=\frac{k_2r_1(\alpha_2 k_1 +r_2)}{r_1r_2-\alpha_1\alpha_2 k_1k_2}$$
We assume that
\begin{equation}\label{conditoin}
r_1r_2-\alpha_1\alpha_2 k_1k_2>0
\end{equation}
and therefore $e_1, e_1>0$.  As a result, we can apply the theocratical results for cooperative systems in \cite{Weinberger2002-1,Wang2010jns} to calculate the minimum speed of (\ref{cooperative}) for information propagation from $(0,0)$ and $(e_1, e_2)$. It was shown in \cite{Weinberger2002-1,Wang2010jns} that there is a traveling wave solution connecting $(0,0)$ and $(e_1, e_2)$ and  the minimum speed of the information propagation can be calculated by the formula (\ref{mini}). For simplicity, assume that and $d_1 \geq d_2$ and $r_1 \geq r_2$. Now it is easy to calculate that the Jacobian of (\ref{cooperative}) at $(0,0)$ is $$
\left(
  \begin{array}{cc}
    r_1 & 0 \\
    0 & r_2 \\
  \end{array}
\right)
$$
For $\lambda \geq 0$, the largest eigenvalue $\Psi(A_{\lambda})$  of the matrix  $$
\left(
  \begin{array}{cc}
    d_1\lambda^2 +r_1 & 0 \\
    0 & d_2\lambda^2 +r_2\\
  \end{array}
\right)
$$
is $d_1\lambda^2 +r_1$.  Therefore
$$
\Phi(\lambda)=\frac{1}{\lambda} \Psi(A_{\lambda})=\inf_{\lambda>0}\frac{d_1\lambda^2+r_1}{\lambda}
$$
In view of (\ref{mini}), a standard calculation shows the propagation speed for (\ref{cooperative}) is
$$
c^*=2\sqrt{d_1r_1}
$$
On the other hand, if $d_2 \geq d_1$ and $r_2 \geq r_1$, the propagation speed for (\ref{cooperative}) is
$$
c^*=2\sqrt{d_2r_2}
$$
This indicates that if (\ref{conditoin}) holds, or the effect of the interaction of the two sources are not too large,  the propagation speed for multiple information sources is largely determined by the more popular source.

\subsubsection{ Propagation Speeds for Competing Information} \label{ComputingSource}
If there are two pieces of information to compete each other to maximize its own influences on online social networks, (\ref{CompetingInfo}) may be used to model the interaction. We assume that all coefficient are positive constants as we focus on its long term behavior.  (\ref{CompetingInfo}) is not a cooperative system.  However, we are interested in a transition process connecting two equilibria $(k_1, 0)$ and $(0, k_2)$ of (\ref{CompetingInfo}).   (\ref{CompetingInfo}) can be brought into a cooperative system by the transforation $v_1=u_1$ and $v_2=k_2-u_2$
\begin{equation}\label{CompetingInfo2}
\begin{split}
&\frac{\partial v_1}{\partial t}=d_1 \frac{\partial^2 v_1}{\partial x^2}+r_1 v_1(1-\frac{v_1}{k_1}) -\alpha_1 v_1 (k_2-v_2)\\
&\frac{\partial v_2}{\partial t}=d_2 \frac{\partial^2 v_2}{\partial x^2}-r_2 (k_2-v_2) \frac{v_2}{k_2} +\alpha_2 v_1 (k_2-v_2)
\end{split}
\end{equation}
The Jacobian of (\ref{CompetingInfo2}) $$
\left(
  \begin{array}{cc}
    r_1-2\frac{r_1v_1}{k_1}-\alpha_1 (k_2-v_2) & a_1v_1 \\
    \alpha_2 (k_2-u_2) & -r_2+2\frac{r_2v_2}{k_2}- \alpha_2 v_1 \\
  \end{array}
\right)
$$
has nonnegative off-diagonal elements for $v_1, v_2 \geq 0$ and $v_2 \leq k_2$.  We assume that
\begin{equation}\label{conditoin1}
r_1> \alpha_1 k_2
\end{equation}
to ensure that the growth of $u_1$ sustains even with the competition from $u_2$. We also assume that $d_1 \geq d_2$, information $u_1$ is not less popular than information $u_2$.

We are interested in a transition process connecting two equilibria $(0,0)$ and $(k_1, k_2)$ of (\ref{CompetingInfo2}) that correspond to the two equilibria $(0, k_2)$ and $(k_1, 0)$ of (\ref{CompetingInfo}). Again we can apply the theocratical results for cooperative systems in \cite{Weinberger2002-1,Wang2010jns} to calculate the minimum speed of (\ref{CompetingInfo2}) for information propagation from $(0,0)$ and $(k_1, k_2)$. It was shown in \cite{Weinberger2002-1,Wang2010jns} that there is a traveling wave solution of (\ref{CompetingInfo2}) connecting its two equilibria $(0,0)$ and $(k_1, K_2)$ and  the minimum speed of the information propagation can be calculated by the formula (\ref{mini}).  Now it is easy to calculate that the Jacobian of (\ref{CompetingInfo2}) at $(0,0)$ is $$
\left(
  \begin{array}{cc}
    r_1-\alpha_1 k_2 & 0 \\
    \alpha_2 k_2 & -r_2 \\
  \end{array}
\right)
$$
For $\lambda \geq 0$, the largest eigenvalue $\Psi(A_{\lambda})$  of the matrix  $$
\left(
  \begin{array}{cc}
    d_1\lambda^2 +r_1-\alpha_1 k_2 & 0 \\
    \alpha_2 k_2 & d_2\lambda^2 - r_2\\
  \end{array}
\right)
$$
is $d_1\lambda^2 +r_1-\alpha_1 k_2$.  Therefore
$$
\Phi(\lambda)=\frac{1}{\lambda} \Psi(A_{\lambda})=\inf_{\lambda>0}\frac{d_1\lambda^2 +r_1-\alpha_1 k_2}{\lambda}
$$
In view of (\ref{mini}), a standard calculation shows the propagation speed for (\ref{CompetingInfo2}) is
$$
c^*=2\sqrt{d_1(r_1-\alpha_1 k_2)}
$$
The conclusion indicates that information $u_1$ will win the competition if the growth of $u_1$ sustains even with the competition from $u_2$, and information $u_1$ is not less popular than information $u_2$.  The propagation speed of the information is largely determined by the popularity and growth of the winner minus negative affect from the competition.

\subsubsection{Propagation Speeds for Spatial Epidemiological Models}  \label{Epide}
(\ref{S-PDE}) is a non cooperative systems. In general it still is an open question to show it is linearly determinate. Nevertheless, for (\ref{S-PDE}), we can show $c^*$ is the cut-off point of the existence of traveling wave solutions. Here we would like to examine the traveling solutions and information adoption rate  of a spatial SIR model for information diffusion in online social networks in a closed population consisting of susceptible individuals ($S(t)$), infected individuals ($I (t)$) (i.e. adopted the information) and removed
individuals ($R(t)$) (i.e. refractory). The diffusive SIR model with the standard incidence takes the following form
\begin{equation}\label{S-PDE}
\begin{split}
  \p_tS&=d_1\p_{xx}S-\b SI/(S+I)\\
  \p_tI&=d_2\p_{xx}I+\b SI/(S+I)-\g I\\
  \p_tR&=d_3\p_{xx}R+\g I
\end{split}
\end{equation}
here $\g$ is the remove rate of the infected group,  $\b$ is the adoption (or influence) rate between the susceptible and infectious groups.  $d_1, d_2, d_3>0$ represents the popularity of information with each of the groups.  In general, the information is more popular for the $ I$ group than the $S$ group and, there for, it is assumed that $d_2 \geq d_1$. For the long-term propagation of information in online social networks, it is understood that the adoption rate $\b$ can be constant. (\ref{S-PDE}) is  an extension of the $SI$ model (\ref{SIS}) with the refractory group $R$. A traveling wave solution  of (\ref{S-PDE}) with the form $(S(x+ct), I(x+ct, t), R(x+ct,t)$ represent the transition process of information diffusion from the initial adoption-free equilibrium $(S_{-\infty}, 0, R_{-\infty})$ to another adoption-free state $(S_\infty,0,R_{\infty})$ with $S_\infty$ being determined by the influence rate $\b$ and the remove rate $\g$, as well as possibly the popularity of information. As such, it is important to determine whether traveling waves exist and what the propagation speed $c$ is. Thus we shall look for traveling wave solutions of the form $(S(x+ct), I(x+ct), R(x+ct))$. Because $R$ does not appear in the system of equations for the susceptible individuals $S$ and infected individuals $I$, we omit the $R$ equation and study the following system with $S$ and $I$ only.

\begin{equation}\label{S-PDE0}
\begin{split}
  \p_tS&=d_1\p_{xx}S-\b SI/(S+I)\\
  \p_tI&=d_2\p_{xx}I+\b SI/(S+I)-\g I
\end{split}
\end{equation}
and satisfy the following boundary conditions at infinity:
\begin{equation}\label{bc}
  S(-\infty)=S_{-\infty},\ S(\infty)<S_{-\infty},\ I(\pm\infty)=0.
\end{equation}
In the context of infectious disease, Wang, the first author and Wu \cite{WangXiangS2012} studied the traveling waves and propagation speed of (\ref{S-PDE0}). The result of \cite{WangXiangS2012} is applicable for information diffusion in online social networks. For (\ref{S-PDE0}), the nonlinearity $\f$ in (\ref{eq11}) is no longer cooperative and some of the off-diagonal elements of $\f'$ may be negative.  It is still an open question what additional conditions would guarantee that $\Phi(\lambda)$ maintains the convex-like property. However, for (\ref{S-PDE0}), $\Phi(\lambda)$  is a convex function and we note that the minimum wave speed can be obtained by its linearization at the initial state $(S_{-\infty},0)$. In fact, it is easy to calculate that the Jacobian of (\ref{S-PDE0}) at $(S_{-\infty},0)$ is $$
\left(
  \begin{array}{cc}
    0 & -\beta \\
    0 & \beta -\gamma \\
  \end{array}
\right)
$$
Its largest eigenvalue is $\beta-\g$. For $\mu \geq 0$ and $d_2 \geq d_1$, the largest eigenvalue $\Psi(A_{\lambda})$  of the matrix  $$
\left(
  \begin{array}{cc}
    d_1\lambda^2 & -\beta \\
    0 & d_2\lambda^2+\beta-\g \\
  \end{array}
\right)
$$
is $d_2\mu^2+\beta-\g$.  Therefore
$$
\Phi(\lambda)=\frac{1}{\lambda} \Psi(A_{\lambda})=\inf_{\lambda>0}\frac{d_2\lambda^2+\beta-\g}{\lambda}
$$
In view of (\ref{mini}), a standard calculation shows the wave speed for (\ref{S-PDE0}) is
$$
c^*=2\sqrt{d_2(\b-\g)}
$$
In addition, \cite{WangXiangS2012} shows that $c^*=2\sqrt{d_2(\b-\g)}$ is the cut-off value of $c$ for which there is a traveling wave for (\ref{S-PDE0}) of the form $(S(x+ct),I(x+ct))$. Specifically, it is shown in \cite{WangXiangS2012} that if $R_0:=\b/\g>1$ ($R_0$ is the basic reproduction number for the corresponding ordinary differential system) and $c>c^*:=2\sqrt{d_2(\b-\g)}$ , then there exists a non-trivial and non-negative traveling wave solutions $(S,I)$  of (\ref{S-PDE0}) such that the boundary conditions (\ref{bc}) are satisfied. Furthermore, $S$ is monotonically decreasing, $0\le I(x)\le S(-\infty)-S(\infty)$ for all $x\in\R$, and
  \begin{equation}\label{finalsize}
  \int_{-\infty}^\infty\g I(x)dx=\int_{-\infty}^\infty{\b S(x)I(x)\over S(x)+I(x)}dx=c[S(-\infty)-S(\infty)].
  \end{equation}
On the other hand, if $R_0=\b/\g\le1$ or $c<c^*:=2\sqrt{d_2(\b-\g)}$, then there exist no non-trivial and non-negative traveling wave solution $(S,I)$  of (\ref{S-PDE0}) satisfying the boundary conditions (\ref{bc}).

$c^*=2\sqrt{d_2(\b-\g)}$ is particularly of  interest as it is the cut-off point for the existence of traveling waves of (\ref{S-PDE0}). In other words, the cut-off speed for traveling waves of (\ref{S-PDE0}) is determined by its linearized systems.  $c^*=2\sqrt{d_2(\b-\g)}$ can be viewed as the speed of (\ref{eq11}) for information to spread in a social network.  The result also indicates that the diffusion speed of information is proportional to the square root of the product of the popularity of the information and difference of the adoption rate and remove rate of the adopted group.

\section{Concluding Remarks}\label{endd}

In this paper, we review the recent development in modeling information diffusion in online social networks with partial differential equations.  Building on intuitive cyber-distance, we propose a number of reaction-diffusion equations to characterize information spreading in temporal and spatial dimensions. We start with a number of simple spatial models with extensive validations from real datasets collected from popular online social networks such as Digg.com and Twitter.com. Our experiment results show that the model achieves high accuracy for the majority of news with more than $3000$ votes in Digg and Twitter. In general, our models can achieve over 90\% accuracy.  We discover strong evidence of the feasibility to model the information diffusion process in online social networks such as Digg and Twitter. We also present a number of spatial models for complex interactions. The PDEs models take into account influences from various out-of-network sources such as the mainstream media, and provide a new analytic framework to study the interplay of structural and topical influences on information diffusion over social media.

To the best of our knowledge, our work is the first attempt to propose PDE-based models for characterizing and predicting the temporal and spatial patterns of information diffusion over online social networks. The temporal and spatial characteristics of information diffusion process sheds light on how information spreads and to what extend external influences affect information diffusion over online social networks.  We are in the process to validate the theoretical results such as news propagation speeds we present in this paper.  Our goal is to predict the information diffusion process for a given news story based on the initial phase of information spreading. Our future works include examining how parameter estimations of the models are related to information contents as there are significant differences in the mechanics of information diffusion across topics. These parameters will provide key measurements to quantify online user interactions in online social networks and therefore can be used to classify news stories in online social networks. Mathematical analysis such as bifurcation analysis of the models plays a significant role in parameter estimations. In addition, mathematical analysis of the PDE model with heterogeneity in distance can shed new light on the identification of influential spreader or opinion leaders in online social networks. Not only the mathematical study of the models further confirm the validity of the models, but also reveal and predict new mechanisms governing information flow in social media.  As we can see from the paper, there is a daunting task to analytically and numerically study mathematical problems arising from social media. The complexity of human interactions and rapid change of social media make PDE models from social media even more complex.  We choose simple, yet accurate PDE models in this paper to highlight the new opportunities and challenges for modeling information diffusion over online social networks for mathematicians as well as computer scientists and researchers in social media.

\renewcommand{\baselinestretch}{1.1}
{\small
{\footnotesize
{\scriptsize
{}

}

\end{document}